\documentclass[superscriptaddress,nofootinbib,twocolumn]{revtex4-1}

\usepackage[utf8]{inputenc}

\usepackage[colorlinks]{hyperref}
\hypersetup{
     colorlinks   = true,
     citecolor    = red,
	linkcolor=blue
}

\usepackage{bbm}
\usepackage{amsfonts}
\usepackage{mathrsfs}

\usepackage{amsmath,amssymb}

\usepackage{epsfig}
\usepackage{graphicx}               
\usepackage{url}
\usepackage{float}
\usepackage{soul}
\usepackage{pstricks}
\usepackage{color}
\usepackage{multirow}

\makeatletter
\def\simgt{\mathrel{\lower2.5pt\vbox{\lineskip=0pt\baselineskip=0pt
           \hbox{$>$}\hbox{$\sim$}}}}
\def\simlt{\mathrel{\lower2.5pt\vbox{\lineskip=0pt\baselineskip=0pt
           \hbox{$<$}\hbox{$\sim$}}}}
\makeatother

\def\mysection#1{{\bf #1.} }

\newcommand{\be}{\begin{equation}}
\newcommand{\ee}{\end{equation}}
\newcommand{\bea}{\begin{eqnarray}}
\newcommand{\eea}{\end{eqnarray}}
\newcommand{\beq}{\begin{eqnarray}}
\newcommand{\eeq}{\end{eqnarray}}

\newcommand{\cell}{V_{\text{cell}}}

\definecolor{cerulean}{rgb}{0., 0.52,0.65}

\def\lsim{\mathrel{\rlap{\lower4pt\hbox{\hskip1pt$\sim$}}
     \raise1pt\hbox{$<$}}}         
\def\gsim{\mathrel{\rlap{\lower4pt\hbox{\hskip1pt$\sim$}}
     \raise1pt\hbox{$>$}}}         

\begin{document}

\title{Detection of Light Dark Matter With Optical Phonons in Polar Materials}

\author{Simon Knapen}
\affiliation{Theoretical Physics Group, Lawrence Berkeley National Laboratory, Berkeley, CA 94720 \\ Berkeley Center for Theoretical Physics, University of California, Berkeley, CA 94720}
\author{Tongyan Lin}
\affiliation{Theoretical Physics Group, Lawrence Berkeley National Laboratory, Berkeley, CA 94720 \\ Berkeley Center for Theoretical Physics, University of California, Berkeley, CA 94720}
\affiliation{Department of Physics, University of California, San Diego, CA 92093, USA }
\author{Matt Pyle}
\affiliation{Department of Physics, University of California, Berkeley, CA 94720}
\author{Kathryn M. Zurek}
\affiliation{Theoretical Physics Group, Lawrence Berkeley National Laboratory, Berkeley, CA 94720 \\ Berkeley Center for Theoretical Physics, University of California, Berkeley, CA 94720}

\begin{abstract}
We show that polar materials are excellent targets for direct detection of sub-GeV dark matter due to the presence of gapped optical phonons as well as acoustic phonons with high sound speed.
We take the example of Gallium Arsenide (GaAs), which has the properties needed for experimental realization, and where many results can be estimated analytically. We find GaAs has excellent reach to dark photon absorption, can completely cover the freeze-in benchmark for scattering via an ultralight dark photon, and is competitive with other proposals to detect sub-MeV dark matter scattering off nuclei.
\end{abstract}

\maketitle

\mysection{Introduction}
\label{sec:intro}
The scope of dark matter (DM) searches in recent years has dramatically broadened beyond traditional candidates such as the weakly interacting massive particle (WIMP) and axion.  Theoretically compelling candidates exist in hidden sectors consisting of DM and new light mediators, with numerous mechanisms for setting the DM relic density.
These models have motivated a suite of new direct detection experiments, aimed at sub-GeV DM.  
SuperCDMS \cite{Agnese:2014aze,Agnese:2015nto,Agnese:2016cpb}, DAMIC \cite{Aguilar-Arevalo:2016ndq}, SENSEI \cite{Tiffenberg:2017aac}, NEWS-G \cite{Arnaud:2017bjh} and CRESST \cite{Angloher:2015ewa} are working to detect energy depositions as small as an eV from scattering of MeV mass DM, or absorption of eV mass DM. There are also proposals for eV-scale detection with {\it e.g.}~atoms~\cite{Essig:2012yx}, graphene~\cite{Hochberg:2016ntt}, liquid helium~\cite{Guo:2013dt}, scintillators~\cite{Derenzo:2016fse}, molecular bonds~\cite{Essig:2016crl}, and crystal defects~\cite{Kadribasic:2017obi,Budnik:2017sbu}.

For DM in the 10 keV$-$GeV mass range, freeze-in DM interacting with an ultralight dark photon~\cite{Hall:2009bx,Bernal:2017kxu,Essig:2011nj,Chu:2011be,Essig:2015cda} or asymmetric dark matter~\cite{Kaplan:2009ag,Petraki:2013wwa,Zurek:2013wia} are compelling candidates.
Freeze-in selects a clear target for the scattering rate, while there is also a wide parameter space of asymmetric DM. Other viable DM candidates below an MeV include DM scattering through a light scalar mediator coupled to nucleons~\cite{Knapen:2017xzo,Green:2017ybv}.  In the meV$-$eV mass range, dark photon DM can be absorbed in the same experiment.

To be sensitive to such light DM, a target must have a sufficiently small gap to excitations, as well as favorable kinematics for DM scattering. The first proposals included detecting sub-MeV DM scattering off electrons with a superconducting target~\cite{Hochberg:2015pha,Hochberg:2015fth}, and off nuclei in superfluid helium~\cite{Knapen:2016cue,Schutz:2016tid}.  In these cases, the sensitivity to DM scattering via an ultralight dark photon was limited due to the strong in-medium screening in superconductors, and due to the limited polarizability in superfluid helium. Dirac materials have an excellent reach for this scenario~\cite{Hochberg:2017wce} but such materials  have not yet been produced in the quantities needed for direct detection.

In this {\emph{Letter}} we argue that polar materials are an excellent target for sub-MeV DM, especially for scattering through an ultralight dark photon mediator. There are four reasons for this:  first, these materials feature gapped optical phonons which can be thought of as oscillating dipoles. These dipoles have a sizable coupling to kinetically mixed dark photons; furthermore, the suppression from screening effects is much smaller than in other materials such as superconductors. Second, optical phonons are gapped excitations with typical energies of $\sim30$ meV up to $\sim 100$ meV. This is kinematically favorable for sub-MeV DM, allowing $\gg$ meV energy depositions with low momentum transfer.  Third, the anisotropy of the crystal  induces a directional dependence in the DM scattering rate. Finally, similar to germanium and silicon, the technology already exists to make ultra pure polar materials in bulk.  

Here we show that GaAs exhibits all of these features, with excellent sensitivity to scattering through dark photon and scalar mediators, as well as to dark photon absorption. Furthermore, GaAs has a relatively simple crystal structure, such that many results can be estimated analytically. In a future paper, we will explore sapphire (Al$_2$O$_3$), where the more complex crystal structure is more suitable for directional detection~\cite{longpaper}.

\begin{figure}[t!]
\hspace{-0.5cm}
\includegraphics[width=0.5\textwidth]{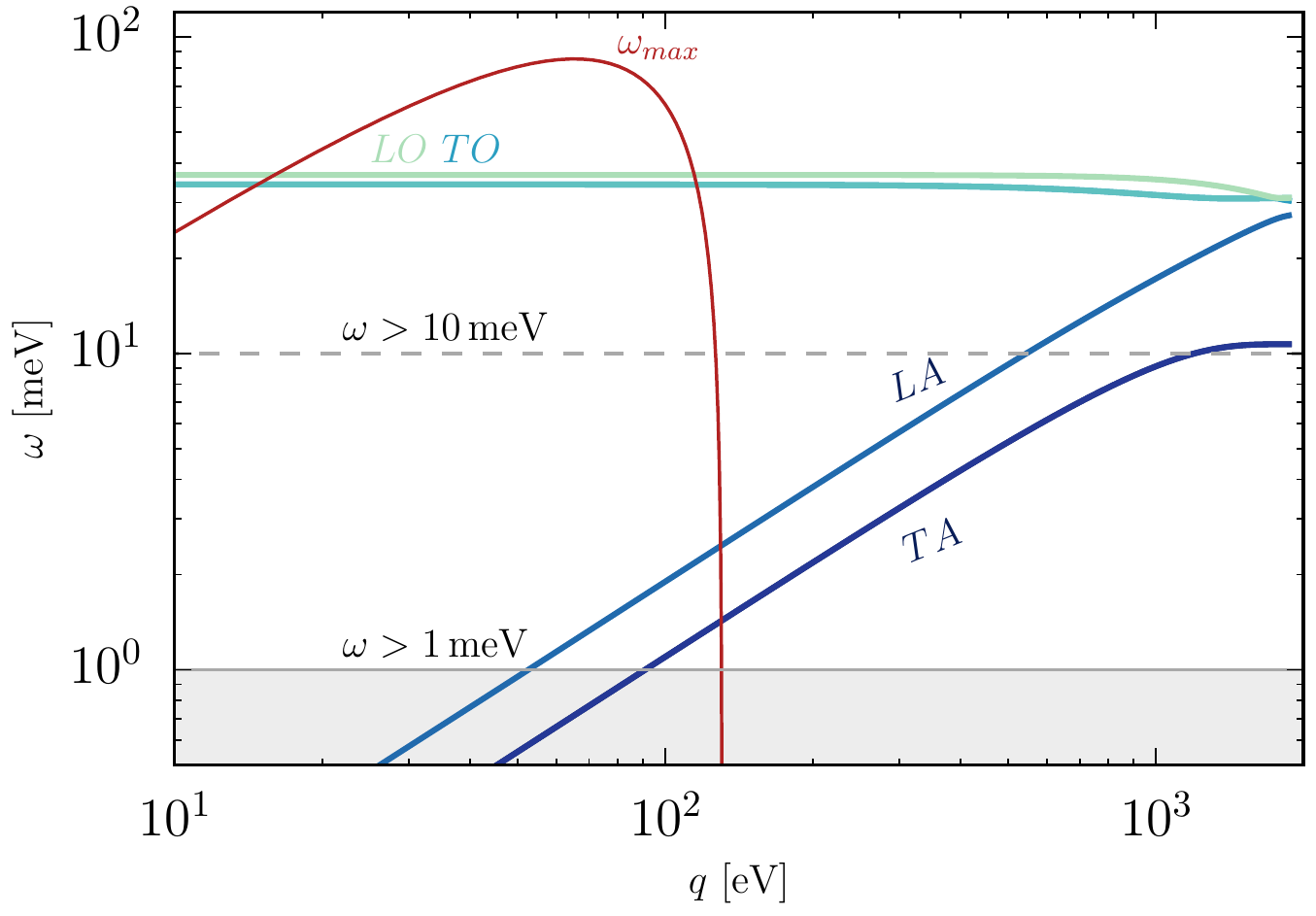}
 \caption{Phonon modes in GaAs for ${\bf q}$ vectors along the direction from $\Gamma = (0,0,0)$ to $X = (0, 2\pi/a, 0) \approx (0,2.21, 0) \, \textrm{keV}$, calculated with \texttt{QuantumESPRESSO} \cite{0953-8984-21-39-395502}. For a representative DM mass $m_X = 25$ keV, we show the maximum energy deposited $\omega_{\rm max}$ as a function of momentum transfer $q$.
Also shown are two possible experimental thresholds, $\omega > 1$ meV and  $\omega > 10$ meV.
 \label{fig:phononmodes}}
\end{figure}

\mysection{Optical Phonons in Polar Materials}
Optical phonons arise when there is more than one atom per primitive unit cell of a crystal. For GaAs, with two atoms in the primitive cell, the phonons consist of two transverse acoustic (TA) modes, one longitudinal acoustic (LA) mode, and similarly two transverse (TO) and one longitudinal (TO) gapped optical modes. Given a model for the effective ion-ion potential, the phonon frequencies are derived by solving a coupled set of differential equations for the ion displacements in the primitive cell (see e.g.~\cite{Kittel}): the acoustic modes have a linear dispersion $\omega \propto q$ as $q\rightarrow 0$, while the optical modes have non-zero frequencies $\omega_{{\rm LO},{\rm TO}}$ as $q\rightarrow 0$.  
The acoustic (optical) modes describe oscillations where the ion displacements are in phase (anti-phase) in the $q \to 0$ limit. 
The dispersions of all phonons in GaAs are shown in Fig.~\ref{fig:phononmodes}; we see that the typical momentum transfers allowed for light DM, in combination with the experimental threshold, greatly reduces the phase space for scattering off acoustic modes but not for optical modes.

The presence of optical phonons is not sufficient for coupling to dark photons. If the  atoms in the unit cell are identical (such as in Si or Ge), there is no net polarization associated with optical phonon oscillations. Instead, in GaAs the ions have net Born effective charges of $\pm2.1$~\cite{YuCardona}, resulting from the polar GaAs bond. The out-of-phase displacements of the optical mode therefore give rise to coherently oscillating dipole moments, which generate long-range dipole fields.  This allows a coupling of the LO phonons to charged particles, including  conduction electrons as well as DM coupled to an ultralight dark photon mediator, where in the latter case the DM effectively carries a tiny electric charge. Combined, the gapped dispersion and the dipole moment for optical phonons are crucial for polar materials to be effective targets for scattering and absorption of light DM. 

The optical phonons also contribute to the optical response for energies below the electron band gap $\omega_g$, which is an important quantity in determining the sensitivity of a target to dark photon interactions. 
For $\omega < \omega_g $,  the  permittivity of GaAs can be written as~\cite{LOCKWOOD2005404}
\begin{align}
 	\hat \epsilon(\omega) = \epsilon_\infty  \frac{\omega_{\rm LO}^2 - \omega^2 + i \omega \gamma_{\rm LO}}{\omega^2_{\rm TO} -\omega^2 +  i \omega \gamma_{\rm TO}},
\end{align}
where $\gamma_{\rm TO, LO}$ are damping parameters and $\epsilon_\infty$ is the contribution of the electrons for $\omega < \omega_g \approx 1$ eV in GaAs. This result can be generalized in a straightforward way to polar materials with more optical phonon branches, by including a product over the different branches. Note that the dielectric function becomes close to zero at $\omega=\omega_{\rm LO}$: this reflects the fact that an LO phonon may be present in a material even without a driving external field~\cite{YuCardona}.

The permittivity determines the screening of electric (and dark photon) fields, with  $\epsilon_0\equiv\hat\epsilon(0)$ the usual dielectric constant. We use measured values of the GaAs phonon frequencies and damping constants at $ T = 4.2$ K~\cite{PSSB:PSSB2221950110}, appropriate for a cryogenic experiment. In general $\hat \epsilon(\omega)$ is an O(1) number,  {\it without} the strong screening that is typically present for free charges. Thus sensitivity to dark photon interactions is achieved due to the possibility of coupling to the polarizability and due to the relatively mild screening.

\vspace{0.15cm}
\mysection{Experimental Concept}
\label{sec:DetConcept}
The success of polar materials for light DM searches requires the development of detection technology that can trigger on 30 meV - 100 meV of vibrational excitations with minimal dark count rate.  Traditional semiconductor and scintillation sensor techniques are not feasible since the energy depositions are below the electron excitation energies.  Likewise, traditional low temperature calorimeters, where phonons are allowed to fully thermalize within the target before measurement in the temperature sensor,  are not practical because the coupling of ${\cal O}$(10~mK) phonons to the electronic system of the thermometer is extremely poor. One would need very large volume and heat capacity thermal sensors, which have large thermal noise \cite{Pyle_15_OptimizedCalorimeters}.  

Consequently, only detector concepts wherein athermal phonon excitations are collected and sensed before thermalization are viable.  One option is to absorb athermal phonons into a few-monolayer thick layer of superfluid He film on the target surface, which leads to evaporation of a He atom with some probability. These evaporated He atoms could then be either absorbed onto the bare surface of a small volume calorimeter (depositing both the kinetic energy and the binding energy of the He atom \cite{Sivani_95PLA_HeAdsorptionCrystal,More_96PRB_PhononAmplifier}) or ionized with large E-fields near a sharp metal tip and subsequently accelerated onto a calorimeter (depositing the total electrostatic potential energy \cite{Maris_17PRL_HeIonization}).

A second possibility is to instrument the surface of a polar absorber with athermal phonon sensors \cite{FRANK1994367,Irwin_95RSI_QET}, which have been employed by CDMS and also proposed for superconducting DM detectors \cite{Hochberg:2015fth}.  High energy phonons produced by DM interactions quickly decay anharmonically into $\mathcal{O}(\mathrm{10^2})$ acoustic phonons with energies around $\mathcal{O}$(meV). At this energy scale, both isotopic scattering and anharmonic decay timescales become long \cite{Tamura_85PRB_AHDC} compared to travel times across the crystal. The athermal phonons are thus either thermalized via surface down-conversion processes or collected by superconducting collection fins; in the latter case they produce quasiparticles which are detected in a small volume (and thus sensitive) Transition Edge Sensor (TES) or Microwave Kinetic Inductance Device (MKID).  

Clean, well-polished crystal surfaces have been shown to have an athermal phonon surface thermalization probability of less than 10$^{-3}$ at 10 mK \cite{Knaak_86_SurfaceThermalization}, so only a small fraction ($< 1\%$) of the total detector surface area must be instrumented to collect nearly all athermal phonons. Conceptually, this allows for O(1-10 meV) sensitivity with a 125 mm$^{3}$ absorber volume as shown in  \cite{Hochberg:2015fth}.   

Radiogenic backgrounds (Comptons, $^3$H, $^{210}$Pb decay products) have typical energy scales that are much larger than the energies of interest here, and thus are not expected to be problematic given demonstrated capabilities for controlling such backgrounds \cite{SCDMS_17PRD_SensitivityProjections}.  The dominant remaining particle backgrounds in such an experiment are $pp$ neutrinos, where  a few events per kg-year can be expected \cite{Hochberg:2015fth}, and coherent scattering of high-energy photons \cite{2016arXiv161007656R}, which we estimate to be $\sim 50$ events/kg-year accounting for structure effects.  The latter background can be suppressed to the $\sim 10^{-2}$ level with an active veto on the hard photon, and so we take the zero background limit for our projections.

\vspace{0.2cm}
\mysection{Dark photon absorption}
\label{sec:DP}
We first consider DM consisting of nonthermally-produced dark photons with kinetic mixing given by $-\kappa F^\prime_{\mu \nu}F^{\mu \nu}/2$, for the mass range of $\approx$ meV - 100 eV. The DM can be detected through absorption, where all of the mass-energy of the DM goes into the excitation.  
The absorption rate can be related to the optical properties of the material (see Ref.~\cite{An:2014twa,Hochberg:2016ajh}):
\beq
R =  \frac{1}{\rho}\frac{\rho_{\rm DM}}{m_{A'}} \kappa_{\rm eff}^2 \sigma_1.
\eeq
where $\sigma_1$ is the absorption rate of photons,  $\rho$ is the mass density of the target and $\rho_{\rm DM} = 0.3 \mbox{ GeV/cm}^3$ is the local DM density. 
$\kappa_{\rm eff}$ is the in-medium coupling of $A'$ with the EM current, obtained by diagonalizing the in-medium polarization tensors for the photon and dark photon:
\beq
\kappa_{\rm eff}^2 = \frac{\kappa^2 m_{A'}^4}{\left[m_{A'}^2 - \mbox{Re}~\Pi(\omega) \right]^2 + \mbox{Im}~\Pi(\omega)^2}.
\eeq
$\Pi(\omega) = -i \sigma \omega$ is the photon polarization tensor in the $\mathbf{q}\rightarrow0$ limit, valid for absorption processes where $| {\bf q}| \ll \omega$. $\sigma$ is the complex optical conductivity. From the optical theorem, the absorption rate is given by the real part of the optical conductivity, $\sigma_1 =  - \frac{{\rm Im} \Pi(\omega)}{\omega}$. Finally, these quantities are related to the permittivity of a material by $\hat \epsilon = \hat n^2 = 1 + i \sigma/\omega$ with $\hat n$ the complex index of refraction.

\begin{figure}[t]
\hspace{-0.5cm}
\includegraphics[width=0.5\textwidth]{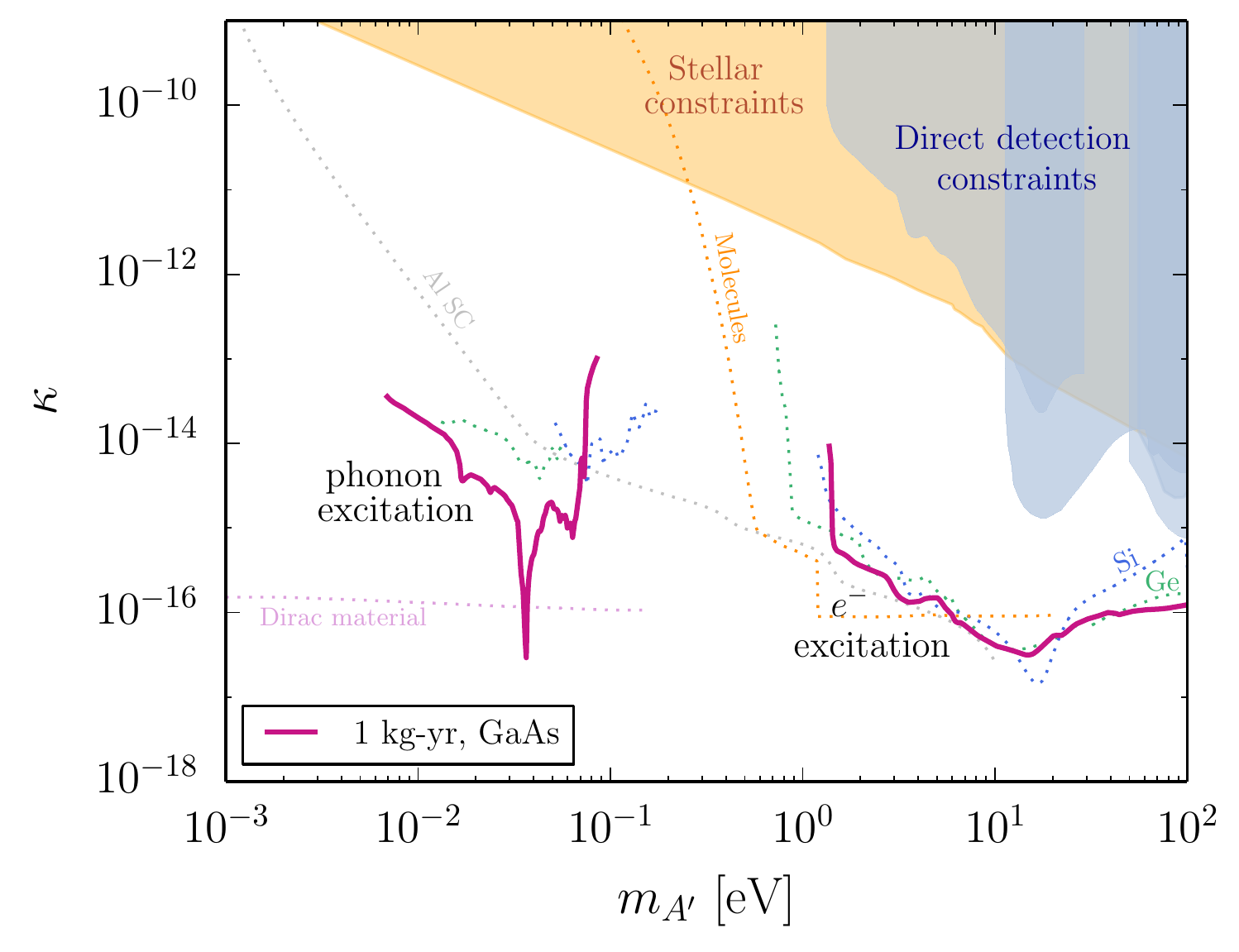}
\vspace{-0.7cm}
 \caption{Reach for absorption of dark photon DM, in terms of the kinetic mixing parameter $\kappa$ for kg-year exposure. Shaded regions are stellar constraints~\cite{An:2013yfc,An:2013yua},  and direct detection constraints from DAMIC~\cite{Aguilar-Arevalo:2016zop},  Xenon10~\cite{An:2014twa,Bloch:2016sjj}, Xenon100~\cite{Hochberg:2016sqx,Bloch:2016sjj}, and CDMSlite~\cite{Hochberg:2016sqx}. The dotted lines are the projected reach with an Al superconductor~\cite{Hochberg:2016ajh}, Ge and Si  semiconductors~\cite{Hochberg:2016sqx}, Dirac materials~\cite{Hochberg:2017wce} and molecules~\cite{Arvanitaki:2017nhi}. See Ref.~\cite{Bloch:2016sjj}  for absorption on GaAs for \mbox{$m_{A'} > $ eV}, and Ref.~\cite{Bunting:2017net} for the reach of molecular magnets.
 \vspace{-0.3cm} \label{fig:darkphotonAbs}}
\end{figure}

To determine the reach on the kinetic mixing parameter $\kappa$, we use calculations of the sub-eV absorption coefficient in the $T=0$ limit from Ref.~\cite{PhysRevB.70.245209}, supplemented with the optical conductivity data of Ref.~\cite{opticalconstantsGaAs} that extends up to 100 eV. 
The result is shown in Fig.~\ref{fig:darkphotonAbs}, assuming 3 events for a kg-year exposure. The reach below $100$ meV is obtained from absorption into phonon modes; there is resonant absorption into the LO phonon at $m_{A'}\approx36$ meV, as well as sidebands from multiphonon processes. The reach for $m_{A'} >$ eV is due to electron excitations above the bandgap, considered before in Ref.~\cite{Bloch:2016sjj}.

\begin{figure}[t!]
\hspace{-0.5cm}
\includegraphics[width=0.5\textwidth]{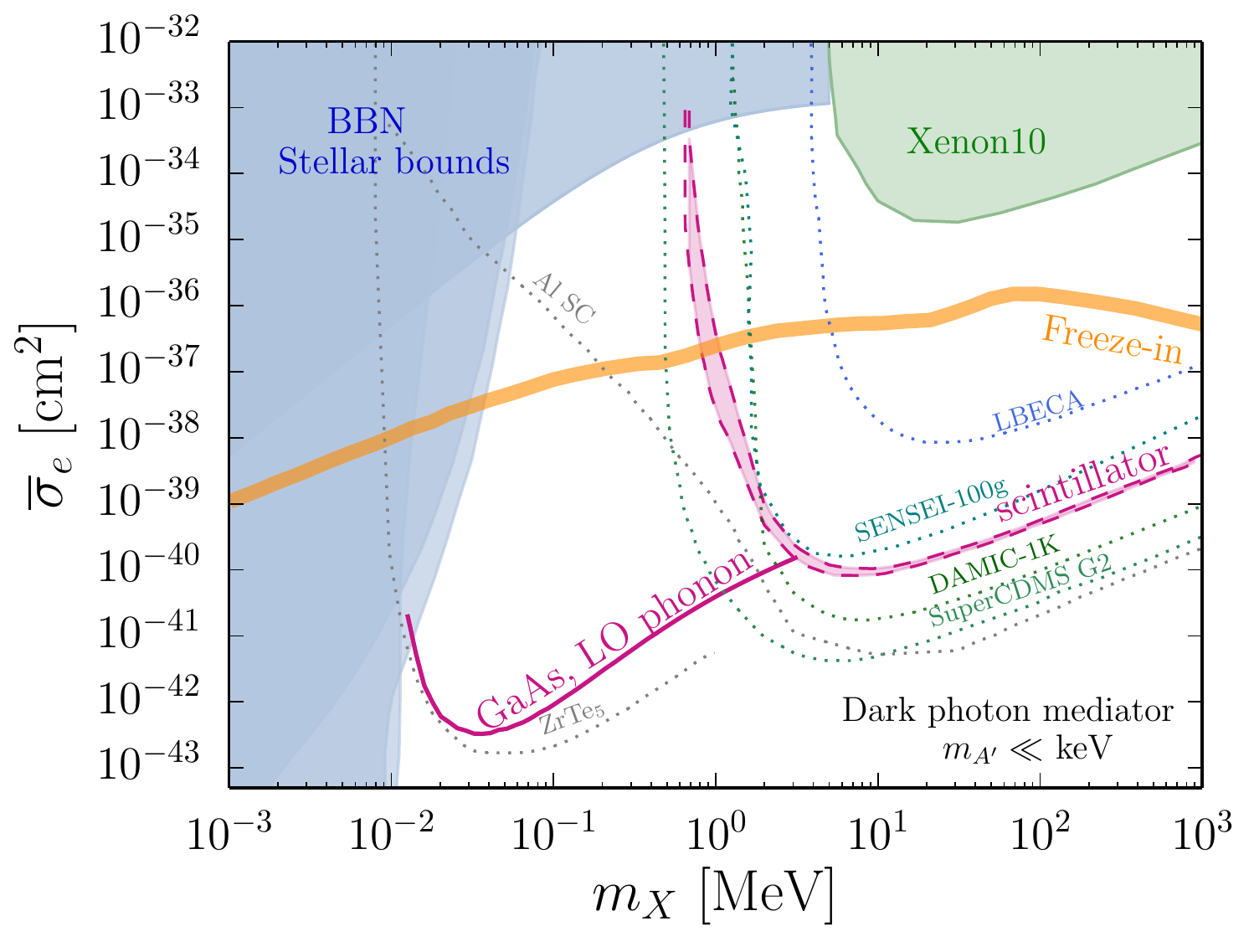}
\vspace{-0.7cm}
 \caption{Sensitivity to DM scattering via an ultralight dark photon, for kg-yr exposure on GaAs. On the orange line the relic abundance can be explained by freeze-in~\cite{Essig:2011nj,Chu:2011be,Essig:2015cda}. The reach for $m_X < $ MeV is from scattering into LO phonons.  For $m_X >$ MeV, the reach comes from considering GaAs as a scintillator for DM-electron scattering~\cite{Derenzo:2016fse}. The blue region indicates stellar~\cite{Vogel:2013raa} and BBN constraints~\cite{Davidson:2000hf}, while the green region is a Xenon10 limit~\cite{Essig:2012yx}.
 Projections for various experimental proposals are from Refs.~\cite{Essig:2015cda,Battaglieri:2017aum,Hochberg:2015pha,Hochberg:2017wce} (dotted lines).
\vspace{-0.3cm} }  \label{fig:DPScatt}
\end{figure}

\vspace{0.15cm}
\mysection{DM scattering via ultralight dark photon} 
In this case we assume a fermionic DM interaction $g_X \overline X \gamma^\mu X A'_\mu$, in addition to kinetic mixing. Taking the limit $m_{A'}\ll$ eV, the results are best understood in the basis where $X$ is effectively millicharged under the standard model photon with coupling $\kappa g_X \overline X \gamma^\mu X A_\mu$ (see {\it e.g.}~appendix D of \cite{Knapen:2017xzo}). The interaction of $X$ with an LO phonon is effectively that of a test charge with electric charge $\kappa g_X$.
We can then follow the derivation of the well-known Fr\"ohlich Hamiltonian for interactions of electrons with LO phonons in the long-wavelength and isotropic limit~\cite{frolich1954,YuCardona,BornHuang,PhysRevLett.115.176401}. These long-range interactions are important in explaining electron mobility data in polar materials, and have previously been computed for GaAs in Refs.~\cite{PhysRevB.94.201201,PhysRevB.92.054307}.
To obtain the interaction of DM with LO phonons in this limit, we rescale the original Fr\"ohlich Hamiltonian by the electric charge ratio of DM to electrons, $ \kappa g_X/e$. This coupling is well-suited to describe scattering of DM in the keV-MeV mass range, with corresponding low momentum transfer $q \lesssim $ keV.
The resulting interaction is
\beq\label{eq:interaction}
	{\cal H}_I = i \frac{\kappa g_X}{e} C_F\sum_{\mathbf{k},\mathbf{q}}  \frac{1}{|\mathbf{q}|}\left[ c_{\bf q}^\dagger a^\dagger_{ \bf k - \bf q} a_{\bf k} - \rm {c.c.}\right]  
\eeq
where $c_{\bf q}^\dagger$  and $a^\dagger_{\bf k}$ are phonon and $X$ creation operators, respectively. The coupling is
\beq
C_F = e \left[ \frac{\omega_{\rm LO}}{2 \cell} \left(\frac{1}{\epsilon_\infty} - \frac{1}{\epsilon_0}\right)\right]^{1/2},
\eeq
where $e$ is the electric charge, $\epsilon_{0}$ ($\epsilon_\infty$) is the static (high frequency) dielectric constant, and $\cell$ is the primitive cell volume. For GaAs, $\epsilon_0=12.9$ and $\epsilon_\infty=10.88$~\cite{LOCKWOOD2005404}.  The above approximations are expected to break down for anisotropic crystals, such as sapphire, and for $m_X\gtrsim1$ MeV. For these DM masses, the typical momentum transfer becomes comparable or larger to the inverse interparticle spacing, requiring a description of processes where phonons are excited outside the first Brillouin zone. In addition, multiphonon processes are expected to contribute and the scattering rate transitions to regular nuclear recoils for sufficiently large momentum and energy deposited. We therefore restrict to the sub-MeV mass regime, while other experimental proposals are well suited for MeV-GeV DM scattering (Fig.~\ref{fig:DPScatt}).

%
Using Eq.~\eqref{eq:interaction}, we find that the scattering rate for $X$ with initial momentum $p_i$ is:
\beq
\Gamma({\bf p_i}) =2 \pi  \int \frac{d^3 {\bf p}_f}{(2 \pi)^3} \delta(E_f - E_i - \omega) |{\cal M}_{\bf q}|^2,
\eeq
with matrix element
\beq
|{\cal M}_{\bf q} |^2 =  \frac{\kappa^2  g_X^2 }{e^2} \frac{C_F^2}{q^2} .
\eeq
The total rate per unit time and target mass is then given by \mbox{$R= \frac{1}{\rho}\frac{\rho_{\rm DM}}{m_X}\int d^3{\bf v} f({\bf v}) \Gamma(m_\chi {\bf v})$}, where $f({\bf v})$ is a boosted, truncated Maxwell-Boltzmann distribution (see e.g.~\cite{Lewin:1995rx}) with velocity dispersion $v_0 = 220$ km/s, Earth velocity $v_e=240$ km/s and escape velocity $v_{esc} = 500$ km/s. To estimate the reach, we require 3 events for a kg-year exposure. As is conventional in the literature, we show in  Fig.~\ref{fig:DPScatt} the resulting sensitivity on $\kappa g_X$ in terms of the DM-electron cross section,
\beq
\bar \sigma_e \equiv \frac{4  \mu_{Xe}^2 \kappa^2 g_X^2 \alpha_{em}}{(\alpha_{em} m_e)^4}. 
\eeq  
where $\alpha_{em}$ is the fine structure constant, $m_e$ is the electron mass, and $\mu_{Xe}$ is the electron-DM reduced mass. We find that even with $\sim$ gram-month exposures, polar materials can reach the freeze-in benchmark. Away from the freeze-in line, a kg-year exposure  can extend the reach of existing proposals by several orders of magnitude.

\vspace{0.15cm}
\mysection{Scalar-mediated nucleon scattering}
\label{sec:NI}
Finally we consider the case of sub-MeV DM with coupling to nucleons only, similar to what was explored in Ref.~\cite{Knapen:2016cue,Schutz:2016tid} for multiphonon production in superfluid helium. The strength of such an interaction can be parametrized by the average DM-nucleon scattering length $\bar b_n$. GaAs improves over helium for several reasons: first, DM can scatter by exciting a single $\sim36$ meV optical phonon, rather than going through higher-order multiphonon interactions. Second, the speed of sound is $\sim 20$ times higher in GaAs, such that the energy of acoustic phonons is higher and better matched to DM kinematics.

The differential DM scattering rate is
\beq
\frac{d^2 \Gamma}{dq d\omega} = \frac{4\pi }{\cell}\frac{q}{ m_X  p_i} S({\bf q},\omega), 
\eeq
where $p_i$ is the initial DM momentum, and $S({\bf q},\omega)$ is the dynamical structure factor, defined in the same way as for neutron scattering.  
In the long-wavelength limit, $S({\bf q},\omega)$ is given by \cite{Schober2014}
\beq
\label{sko}
S({\bf q},\omega)=\frac{1}{2} \sum\limits_{\nu}\frac{|F_{\nu}({\bf q})|^2}{\omega_{\nu,{\bf q}}}\delta(\omega_{\nu,{\bf q}}\!-\!\omega)
\eeq
where $\nu$ sums over the various phonon branches.
The phonon form factor is 
\beq
|F_\nu({\bf q})|^2=\left|\sum_d \frac{\bar b_d}{\sqrt{m_d}}e^{-W_d({\bf q})}{\bf q}\cdot\mathbf{e}_{\nu,d,{\bf q}}e^{-i{\bf q}\cdot \mathbf{r}_d}\right|^2
\eeq
where $d$ labels atoms in the primitive cell with mass $m_d$ and position ${\bf r}_d$. $\bar b_d$ is the scattering length, ${\bf e}_{\nu,d,{\bf q}}$ is the phonon eigenvector of branch $\nu$ and atom $d$ at momentum ${\bf q}$, and $W_d$ the Debye-Waller factor of atom $d$. 

\begin{figure}[t]
\hspace{-0.5cm}
\includegraphics[width=0.5\textwidth]{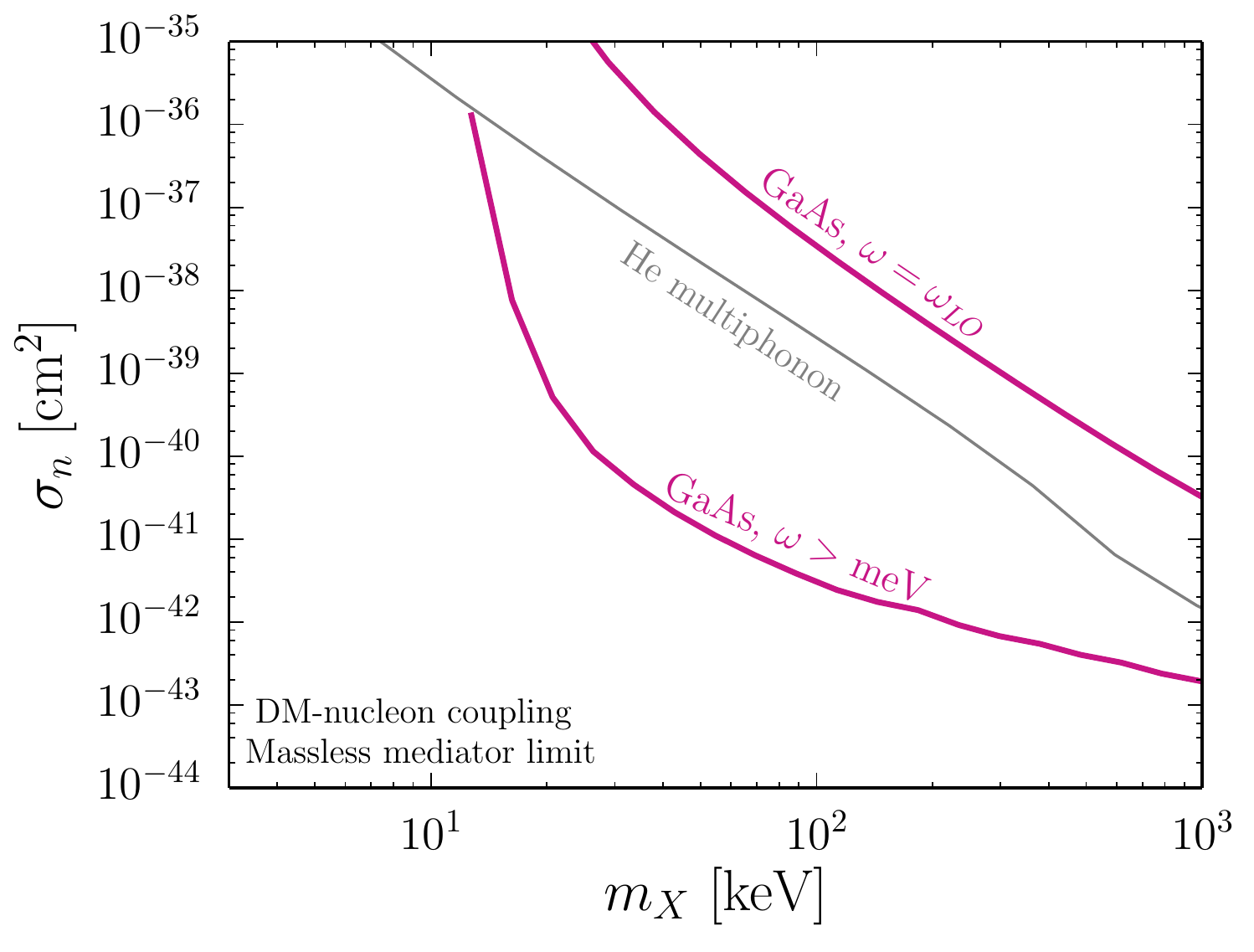}
\vspace{-0.7cm}
 \caption{Sensitivity of GaAs to scattering off nucleons via a scalar mediator, with kg-yr exposure. We consider the projected reach due to production of LO phonons ($\omega = \omega_{\rm LO} \approx $ 36 meV) and that due to production into LA phonons as well, with an even lower threshold $\omega > $ meV.  Also shown is the reach from multiphonon production in superfluid helium~\cite{Knapen:2016cue}. 
\vspace{-0.3cm} }  \label{fig:NSReach}
\end{figure}

Here we estimate the rate in the isotropic and long-wavelength limit where $W_d \approx0$ and the phonon eigenvectors have a simple form:
\beq\label{phononformfact}
	|F_\nu({\bf q})|^2\approx\frac{\bar b^2_n}{2m_n}q^2\left|\sqrt{A_{\text{Ga}}}e^{i{\bf r}_{\rm Ga}\cdot{\bf q}}\pm \sqrt{A_{\text{As}}}e^{i{\bf r}_{\rm As}\cdot{\bf q}}\right|^2
\eeq
with $m_n$ the nucleon mass, $\bar b_n$ the DM-nucleon scattering length and $A_{\text{Ga}}$ ($A_{\text{As}}$) the mass number of Ga (As). The $+$ $(-)$ sign applies to the LA (LO) branch, where both atoms are in phase (anti-phase). For a rough estimate when $m_X \ll$ MeV, the phase factors in \eqref{phononformfact} can be neglected. Similar to the Fr\"ohlich Hamiltonian, the analytic approximations made here are only valid in the sub-MeV mass regime; for larger masses, a reliable theoretical treatment requires a complete description of the phonon band structure over the Brillouin Zone as well as multiphonon processes, which are beyond the scope of this work.

The approximations made here are expected to break down for $m_X\gtrsim1$ MeV. For such masses, the typical momentum transfer becomes comparable to or larger than the inverse interparticle spacing, requiring a description of processes where phonons are excited outside the first Brillouin Zone. In addition, multiphonon processes will contribute and the scattering rate transitions to regular nuclear recoils for sufficiently large momentum and energy deposited. We therefore restrict to the sub-MeV mass regime, while other experimental proposals are well suited for MeV-GeV DM scattering (Fig.~\ref{fig:DPScatt}).  

For scattering via a massless mediator, we also include a $(m_X v_0/q)^4$ form factor and express the  reach in terms of the cross section per nucleon at a reference $q_{\rm ref} = m_X v_0$, $\sigma_n \equiv 4\pi [ \bar  b_n(q_{\rm ref}) ]^2$. The result is shown in Fig.~\ref{fig:NSReach}, where we find a competitive reach with superfluid helium.  The astrophysical and cosmological constraints on this scenario are rather tight but model dependent and hence not shown; see Refs.~\cite{Green:2017ybv,Knapen:2017xzo} for details. The large difference in sensitivity for the optical and acoustic modes is due to the near cancellation in \eqref{phononformfact} for the optical modes, since $A_{\text{Ga}}\approx A_{\text{As}}$. The phase factor in \eqref{phononformfact} also induces a directional dependence for producing optical phonons, which we will explore in future work~\cite{longpaper}.

\vspace{0.2cm}
{\em Acknowledgments.}
We thank Feliciano Giustino, Sinead Griffin, Harikrishnan Ramani, and Dan McKinsey for useful discussions and Sinead Griffin for comments on the manuscript and collaboration on future work.
SK and KZ are supported by the DoE under contract DE-AC02-05CH11231, and SK is also supported in part by the National Science Foundation (NSF) under grants No. PHY-1316783 and No. PHY-1002399. This work was performed in part at the Aspen Center for Physics, which is supported by National Science Foundation grant PHY-1607611 and used resources of the National Energy Research Scientific Computing Center, which is supported by the Office of Science of the DoE under Contract No.~DE-AC02-05CH11231.

\bibliography{sapphire.bib,lightdm.bib}

\begin{thebibliography}{66}%
\makeatletter
\providecommand \@ifxundefined [1]{%
 \@ifx{#1\undefined}
}%
\providecommand \@ifnum [1]{%
 \ifnum #1\expandafter \@firstoftwo
 \else \expandafter \@secondoftwo
 \fi
}%
\providecommand \@ifx [1]{%
 \ifx #1\expandafter \@firstoftwo
 \else \expandafter \@secondoftwo
 \fi
}%
\providecommand \natexlab [1]{#1}%
\providecommand \enquote  [1]{``#1''}%
\providecommand \bibnamefont  [1]{#1}%
\providecommand \bibfnamefont [1]{#1}%
\providecommand \citenamefont [1]{#1}%
\providecommand \href@noop [0]{\@secondoftwo}%
\providecommand \href [0]{\begingroup \@sanitize@url \@href}%
\providecommand \@href[1]{\@@startlink{#1}\@@href}%
\providecommand \@@href[1]{\endgroup#1\@@endlink}%
\providecommand \@sanitize@url [0]{\catcode `\\12\catcode `\$12\catcode
  `\&12\catcode `\#12\catcode `\^12\catcode `\_12\catcode `\%12\relax}%
\providecommand \@@startlink[1]{}%
\providecommand \@@endlink[0]{}%
\providecommand \url  [0]{\begingroup\@sanitize@url \@url }%
\providecommand \@url [1]{\endgroup\@href {#1}{\urlprefix }}%
\providecommand \urlprefix  [0]{URL }%
\providecommand \Eprint [0]{\href }%
\providecommand \doibase [0]{http://dx.doi.org/}%
\providecommand \selectlanguage [0]{\@gobble}%
\providecommand \bibinfo  [0]{\@secondoftwo}%
\providecommand \bibfield  [0]{\@secondoftwo}%
\providecommand \translation [1]{[#1]}%
\providecommand \BibitemOpen [0]{}%
\providecommand \bibitemStop [0]{}%
\providecommand \bibitemNoStop [0]{.\EOS\space}%
\providecommand \EOS [0]{\spacefactor3000\relax}%
\providecommand \BibitemShut  [1]{\csname bibitem#1\endcsname}%
\let\auto@bib@innerbib\@empty
\bibitem [{\citenamefont {Agnese}\ \emph {et~al.}(2014)\citenamefont {Agnese}
  \emph {et~al.}}]{Agnese:2014aze}%
  \BibitemOpen
  \bibfield  {author} {\bibinfo {author} {\bibfnamefont {R.}~\bibnamefont
  {Agnese}} \emph {et~al.} (\bibinfo {collaboration} {SuperCDMS}),\ }\href
  {\doibase 10.1103/PhysRevLett.112.241302} {\bibfield  {journal} {\bibinfo
  {journal} {Phys. Rev. Lett.}\ }\textbf {\bibinfo {volume} {112}},\ \bibinfo
  {pages} {241302} (\bibinfo {year} {2014})},\ \Eprint
  {http://arxiv.org/abs/1402.7137} {arXiv:1402.7137 [hep-ex]} \BibitemShut
  {NoStop}%
\bibitem [{\citenamefont {Agnese}\ \emph {et~al.}(2016)\citenamefont {Agnese}
  \emph {et~al.}}]{Agnese:2015nto}%
  \BibitemOpen
  \bibfield  {author} {\bibinfo {author} {\bibfnamefont {R.}~\bibnamefont
  {Agnese}} \emph {et~al.} (\bibinfo {collaboration} {SuperCDMS}),\ }\href
  {\doibase 10.1103/PhysRevLett.116.071301} {\bibfield  {journal} {\bibinfo
  {journal} {Phys. Rev. Lett.}\ }\textbf {\bibinfo {volume} {116}},\ \bibinfo
  {pages} {071301} (\bibinfo {year} {2016})},\ \Eprint
  {http://arxiv.org/abs/1509.02448} {arXiv:1509.02448 [astro-ph.CO]}
  \BibitemShut {NoStop}%
\bibitem [{\citenamefont {Agnese}\ \emph
  {et~al.}(2017{\natexlab{a}})\citenamefont {Agnese} \emph
  {et~al.}}]{Agnese:2016cpb}%
  \BibitemOpen
  \bibfield  {author} {\bibinfo {author} {\bibfnamefont {R.}~\bibnamefont
  {Agnese}} \emph {et~al.} (\bibinfo {collaboration} {SuperCDMS}),\ }\href
  {\doibase 10.1103/PhysRevD.95.082002} {\bibfield  {journal} {\bibinfo
  {journal} {Phys. Rev.}\ }\textbf {\bibinfo {volume} {D95}},\ \bibinfo {pages}
  {082002} (\bibinfo {year} {2017}{\natexlab{a}})},\ \Eprint
  {http://arxiv.org/abs/1610.00006} {arXiv:1610.00006 [physics.ins-det]}
  \BibitemShut {NoStop}%
\bibitem [{\citenamefont {Aguilar-Arevalo}\ \emph {et~al.}(2016)\citenamefont
  {Aguilar-Arevalo} \emph {et~al.}}]{Aguilar-Arevalo:2016ndq}%
  \BibitemOpen
  \bibfield  {author} {\bibinfo {author} {\bibfnamefont {A.}~\bibnamefont
  {Aguilar-Arevalo}} \emph {et~al.} (\bibinfo {collaboration} {DAMIC}),\ }\href
  {\doibase 10.1103/PhysRevD.94.082006} {\bibfield  {journal} {\bibinfo
  {journal} {Phys. Rev.}\ }\textbf {\bibinfo {volume} {D94}},\ \bibinfo {pages}
  {082006} (\bibinfo {year} {2016})},\ \Eprint
  {http://arxiv.org/abs/1607.07410} {arXiv:1607.07410 [astro-ph.CO]}
  \BibitemShut {NoStop}%
\bibitem [{\citenamefont {Tiffenberg}\ \emph {et~al.}(2017)\citenamefont
  {Tiffenberg}, \citenamefont {Sofo-Haro}, \citenamefont {Drlica-Wagner},
  \citenamefont {Essig}, \citenamefont {Guardincerri}, \citenamefont {Holland},
  \citenamefont {Volansky},\ and\ \citenamefont {Yu}}]{Tiffenberg:2017aac}%
  \BibitemOpen
  \bibfield  {author} {\bibinfo {author} {\bibfnamefont {J.}~\bibnamefont
  {Tiffenberg}}, \bibinfo {author} {\bibfnamefont {M.}~\bibnamefont
  {Sofo-Haro}}, \bibinfo {author} {\bibfnamefont {A.}~\bibnamefont
  {Drlica-Wagner}}, \bibinfo {author} {\bibfnamefont {R.}~\bibnamefont
  {Essig}}, \bibinfo {author} {\bibfnamefont {Y.}~\bibnamefont {Guardincerri}},
  \bibinfo {author} {\bibfnamefont {S.}~\bibnamefont {Holland}}, \bibinfo
  {author} {\bibfnamefont {T.}~\bibnamefont {Volansky}}, \ and\ \bibinfo
  {author} {\bibfnamefont {T.-T.}\ \bibnamefont {Yu}},\ }\href {\doibase
  10.1103/PhysRevLett.119.131802} {\bibfield  {journal} {\bibinfo  {journal}
  {Phys. Rev. Lett.}\ }\textbf {\bibinfo {volume} {119}},\ \bibinfo {pages}
  {131802} (\bibinfo {year} {2017})},\ \Eprint
  {http://arxiv.org/abs/1706.00028} {arXiv:1706.00028 [physics.ins-det]}
  \BibitemShut {NoStop}%
\bibitem [{\citenamefont {Arnaud}\ \emph {et~al.}(2018)\citenamefont {Arnaud}
  \emph {et~al.}}]{Arnaud:2017bjh}%
  \BibitemOpen
  \bibfield  {author} {\bibinfo {author} {\bibfnamefont {Q.}~\bibnamefont
  {Arnaud}} \emph {et~al.} (\bibinfo {collaboration} {NEWS-G}),\ }\href
  {\doibase 10.1016/j.astropartphys.2017.10.009} {\bibfield  {journal}
  {\bibinfo  {journal} {Astropart. Phys.}\ }\textbf {\bibinfo {volume} {97}},\
  \bibinfo {pages} {54} (\bibinfo {year} {2018})},\ \Eprint
  {http://arxiv.org/abs/1706.04934} {arXiv:1706.04934 [astro-ph.IM]}
  \BibitemShut {NoStop}%
\bibitem [{\citenamefont {Angloher}\ \emph {et~al.}(2016)\citenamefont
  {Angloher} \emph {et~al.}}]{Angloher:2015ewa}%
  \BibitemOpen
  \bibfield  {author} {\bibinfo {author} {\bibfnamefont {G.}~\bibnamefont
  {Angloher}} \emph {et~al.} (\bibinfo {collaboration} {CRESST}),\ }\href
  {\doibase 10.1140/epjc/s10052-016-3877-3} {\bibfield  {journal} {\bibinfo
  {journal} {Eur. Phys. J.}\ }\textbf {\bibinfo {volume} {C76}},\ \bibinfo
  {pages} {25} (\bibinfo {year} {2016})},\ \Eprint
  {http://arxiv.org/abs/1509.01515} {arXiv:1509.01515 [astro-ph.CO]}
  \BibitemShut {NoStop}%
\bibitem [{\citenamefont {Essig}\ \emph
  {et~al.}(2012{\natexlab{a}})\citenamefont {Essig}, \citenamefont
  {Manalaysay}, \citenamefont {Mardon}, \citenamefont {Sorensen},\ and\
  \citenamefont {Volansky}}]{Essig:2012yx}%
  \BibitemOpen
  \bibfield  {author} {\bibinfo {author} {\bibfnamefont {R.}~\bibnamefont
  {Essig}}, \bibinfo {author} {\bibfnamefont {A.}~\bibnamefont {Manalaysay}},
  \bibinfo {author} {\bibfnamefont {J.}~\bibnamefont {Mardon}}, \bibinfo
  {author} {\bibfnamefont {P.}~\bibnamefont {Sorensen}}, \ and\ \bibinfo
  {author} {\bibfnamefont {T.}~\bibnamefont {Volansky}},\ }\href {\doibase
  10.1103/PhysRevLett.109.021301} {\bibfield  {journal} {\bibinfo  {journal}
  {Phys. Rev. Lett.}\ }\textbf {\bibinfo {volume} {109}},\ \bibinfo {pages}
  {021301} (\bibinfo {year} {2012}{\natexlab{a}})},\ \Eprint
  {http://arxiv.org/abs/1206.2644} {arXiv:1206.2644 [astro-ph.CO]} \BibitemShut
  {NoStop}%
\bibitem [{\citenamefont {Hochberg}\ \emph
  {et~al.}(2016{\natexlab{a}})\citenamefont {Hochberg}, \citenamefont {Kahn},
  \citenamefont {Lisanti}, \citenamefont {Tully},\ and\ \citenamefont
  {Zurek}}]{Hochberg:2016ntt}%
  \BibitemOpen
  \bibfield  {author} {\bibinfo {author} {\bibfnamefont {Y.}~\bibnamefont
  {Hochberg}}, \bibinfo {author} {\bibfnamefont {Y.}~\bibnamefont {Kahn}},
  \bibinfo {author} {\bibfnamefont {M.}~\bibnamefont {Lisanti}}, \bibinfo
  {author} {\bibfnamefont {C.~G.}\ \bibnamefont {Tully}}, \ and\ \bibinfo
  {author} {\bibfnamefont {K.~M.}\ \bibnamefont {Zurek}},\ }\href@noop {} {\
  (\bibinfo {year} {2016}{\natexlab{a}})},\ \Eprint
  {http://arxiv.org/abs/1606.08849} {arXiv:1606.08849 [hep-ph]} \BibitemShut
  {NoStop}%
\bibitem [{\citenamefont {Guo}\ and\ \citenamefont
  {McKinsey}(2013)}]{Guo:2013dt}%
  \BibitemOpen
  \bibfield  {author} {\bibinfo {author} {\bibfnamefont {W.}~\bibnamefont
  {Guo}}\ and\ \bibinfo {author} {\bibfnamefont {D.~N.}\ \bibnamefont
  {McKinsey}},\ }\href {\doibase 10.1103/PhysRevD.87.115001} {\bibfield
  {journal} {\bibinfo  {journal} {Phys. Rev.}\ }\textbf {\bibinfo {volume}
  {D87}},\ \bibinfo {pages} {115001} (\bibinfo {year} {2013})},\ \Eprint
  {http://arxiv.org/abs/1302.0534} {arXiv:1302.0534 [astro-ph.IM]} \BibitemShut
  {NoStop}%
\bibitem [{\citenamefont {Derenzo}\ \emph {et~al.}(2017)\citenamefont
  {Derenzo}, \citenamefont {Essig}, \citenamefont {Massari}, \citenamefont
  {Soto},\ and\ \citenamefont {Yu}}]{Derenzo:2016fse}%
  \BibitemOpen
  \bibfield  {author} {\bibinfo {author} {\bibfnamefont {S.}~\bibnamefont
  {Derenzo}}, \bibinfo {author} {\bibfnamefont {R.}~\bibnamefont {Essig}},
  \bibinfo {author} {\bibfnamefont {A.}~\bibnamefont {Massari}}, \bibinfo
  {author} {\bibfnamefont {A.}~\bibnamefont {Soto}}, \ and\ \bibinfo {author}
  {\bibfnamefont {T.-T.}\ \bibnamefont {Yu}},\ }\href {\doibase
  10.1103/PhysRevD.96.016026} {\bibfield  {journal} {\bibinfo  {journal} {Phys.
  Rev.}\ }\textbf {\bibinfo {volume} {D96}},\ \bibinfo {pages} {016026}
  (\bibinfo {year} {2017})},\ \Eprint {http://arxiv.org/abs/1607.01009}
  {arXiv:1607.01009 [hep-ph]} \BibitemShut {NoStop}%
\bibitem [{\citenamefont {Essig}\ \emph {et~al.}(2017)\citenamefont {Essig},
  \citenamefont {Mardon}, \citenamefont {Slone},\ and\ \citenamefont
  {Volansky}}]{Essig:2016crl}%
  \BibitemOpen
  \bibfield  {author} {\bibinfo {author} {\bibfnamefont {R.}~\bibnamefont
  {Essig}}, \bibinfo {author} {\bibfnamefont {J.}~\bibnamefont {Mardon}},
  \bibinfo {author} {\bibfnamefont {O.}~\bibnamefont {Slone}}, \ and\ \bibinfo
  {author} {\bibfnamefont {T.}~\bibnamefont {Volansky}},\ }\href {\doibase
  10.1103/PhysRevD.95.056011} {\bibfield  {journal} {\bibinfo  {journal} {Phys.
  Rev.}\ }\textbf {\bibinfo {volume} {D95}},\ \bibinfo {pages} {056011}
  (\bibinfo {year} {2017})},\ \Eprint {http://arxiv.org/abs/1608.02940}
  {arXiv:1608.02940 [hep-ph]} \BibitemShut {NoStop}%
\bibitem [{\citenamefont {Kadribasic}\ \emph {et~al.}(2017)\citenamefont
  {Kadribasic}, \citenamefont {Mirabolfathi}, \citenamefont {Nordlund},
  \citenamefont {Holmström},\ and\ \citenamefont
  {Djurabekova}}]{Kadribasic:2017obi}%
  \BibitemOpen
  \bibfield  {author} {\bibinfo {author} {\bibfnamefont {F.}~\bibnamefont
  {Kadribasic}}, \bibinfo {author} {\bibfnamefont {N.}~\bibnamefont
  {Mirabolfathi}}, \bibinfo {author} {\bibfnamefont {K.}~\bibnamefont
  {Nordlund}}, \bibinfo {author} {\bibfnamefont {E.}~\bibnamefont
  {Holmström}}, \ and\ \bibinfo {author} {\bibfnamefont {F.}~\bibnamefont
  {Djurabekova}},\ }\href@noop {} {\  (\bibinfo {year} {2017})},\ \Eprint
  {http://arxiv.org/abs/1703.05371} {arXiv:1703.05371 [physics.ins-det]}
  \BibitemShut {NoStop}%
\bibitem [{\citenamefont {Budnik}\ \emph {et~al.}(2017)\citenamefont {Budnik},
  \citenamefont {Chesnovsky}, \citenamefont {Slone},\ and\ \citenamefont
  {Volansky}}]{Budnik:2017sbu}%
  \BibitemOpen
  \bibfield  {author} {\bibinfo {author} {\bibfnamefont {R.}~\bibnamefont
  {Budnik}}, \bibinfo {author} {\bibfnamefont {O.}~\bibnamefont {Chesnovsky}},
  \bibinfo {author} {\bibfnamefont {O.}~\bibnamefont {Slone}}, \ and\ \bibinfo
  {author} {\bibfnamefont {T.}~\bibnamefont {Volansky}},\ }\href@noop {} {\
  (\bibinfo {year} {2017})},\ \Eprint {http://arxiv.org/abs/1705.03016}
  {arXiv:1705.03016 [hep-ph]} \BibitemShut {NoStop}%
\bibitem [{\citenamefont {Hall}\ \emph {et~al.}(2010)\citenamefont {Hall},
  \citenamefont {Jedamzik}, \citenamefont {March-Russell},\ and\ \citenamefont
  {West}}]{Hall:2009bx}%
  \BibitemOpen
  \bibfield  {author} {\bibinfo {author} {\bibfnamefont {L.~J.}\ \bibnamefont
  {Hall}}, \bibinfo {author} {\bibfnamefont {K.}~\bibnamefont {Jedamzik}},
  \bibinfo {author} {\bibfnamefont {J.}~\bibnamefont {March-Russell}}, \ and\
  \bibinfo {author} {\bibfnamefont {S.~M.}\ \bibnamefont {West}},\ }\href
  {\doibase 10.1007/JHEP03(2010)080} {\bibfield  {journal} {\bibinfo  {journal}
  {JHEP}\ }\textbf {\bibinfo {volume} {03}},\ \bibinfo {pages} {080} (\bibinfo
  {year} {2010})},\ \Eprint {http://arxiv.org/abs/0911.1120} {arXiv:0911.1120
  [hep-ph]} \BibitemShut {NoStop}%
\bibitem [{\citenamefont {Bernal}\ \emph {et~al.}(2017)\citenamefont {Bernal},
  \citenamefont {Heikinheimo}, \citenamefont {Tenkanen}, \citenamefont
  {Tuominen},\ and\ \citenamefont {Vaskonen}}]{Bernal:2017kxu}%
  \BibitemOpen
  \bibfield  {author} {\bibinfo {author} {\bibfnamefont {N.}~\bibnamefont
  {Bernal}}, \bibinfo {author} {\bibfnamefont {M.}~\bibnamefont {Heikinheimo}},
  \bibinfo {author} {\bibfnamefont {T.}~\bibnamefont {Tenkanen}}, \bibinfo
  {author} {\bibfnamefont {K.}~\bibnamefont {Tuominen}}, \ and\ \bibinfo
  {author} {\bibfnamefont {V.}~\bibnamefont {Vaskonen}},\ }\href@noop {} {\
  (\bibinfo {year} {2017})},\ \Eprint {http://arxiv.org/abs/1706.07442}
  {arXiv:1706.07442 [hep-ph]} \BibitemShut {NoStop}%
\bibitem [{\citenamefont {Essig}\ \emph
  {et~al.}(2012{\natexlab{b}})\citenamefont {Essig}, \citenamefont {Mardon},\
  and\ \citenamefont {Volansky}}]{Essig:2011nj}%
  \BibitemOpen
  \bibfield  {author} {\bibinfo {author} {\bibfnamefont {R.}~\bibnamefont
  {Essig}}, \bibinfo {author} {\bibfnamefont {J.}~\bibnamefont {Mardon}}, \
  and\ \bibinfo {author} {\bibfnamefont {T.}~\bibnamefont {Volansky}},\ }\href
  {\doibase 10.1103/PhysRevD.85.076007} {\bibfield  {journal} {\bibinfo
  {journal} {Phys. Rev.}\ }\textbf {\bibinfo {volume} {D85}},\ \bibinfo {pages}
  {076007} (\bibinfo {year} {2012}{\natexlab{b}})},\ \Eprint
  {http://arxiv.org/abs/1108.5383} {arXiv:1108.5383 [hep-ph]} \BibitemShut
  {NoStop}%
\bibitem [{\citenamefont {Chu}\ \emph {et~al.}(2012)\citenamefont {Chu},
  \citenamefont {Hambye},\ and\ \citenamefont {Tytgat}}]{Chu:2011be}%
  \BibitemOpen
  \bibfield  {author} {\bibinfo {author} {\bibfnamefont {X.}~\bibnamefont
  {Chu}}, \bibinfo {author} {\bibfnamefont {T.}~\bibnamefont {Hambye}}, \ and\
  \bibinfo {author} {\bibfnamefont {M.~H.~G.}\ \bibnamefont {Tytgat}},\ }\href
  {\doibase 10.1088/1475-7516/2012/05/034} {\bibfield  {journal} {\bibinfo
  {journal} {JCAP}\ }\textbf {\bibinfo {volume} {1205}},\ \bibinfo {pages}
  {034} (\bibinfo {year} {2012})},\ \Eprint {http://arxiv.org/abs/1112.0493}
  {arXiv:1112.0493 [hep-ph]} \BibitemShut {NoStop}%
\bibitem [{\citenamefont {Essig}\ \emph {et~al.}(2015)\citenamefont {Essig},
  \citenamefont {Fernandez-Serra}, \citenamefont {Mardon}, \citenamefont
  {Soto}, \citenamefont {Volansky},\ and\ \citenamefont {Yu}}]{Essig:2015cda}%
  \BibitemOpen
  \bibfield  {author} {\bibinfo {author} {\bibfnamefont {R.}~\bibnamefont
  {Essig}}, \bibinfo {author} {\bibfnamefont {M.}~\bibnamefont
  {Fernandez-Serra}}, \bibinfo {author} {\bibfnamefont {J.}~\bibnamefont
  {Mardon}}, \bibinfo {author} {\bibfnamefont {A.}~\bibnamefont {Soto}},
  \bibinfo {author} {\bibfnamefont {T.}~\bibnamefont {Volansky}}, \ and\
  \bibinfo {author} {\bibfnamefont {T.-T.}\ \bibnamefont {Yu}},\ }\href@noop {}
  {\  (\bibinfo {year} {2015})},\ \Eprint {http://arxiv.org/abs/1509.01598}
  {arXiv:1509.01598 [hep-ph]} \BibitemShut {NoStop}%
\bibitem [{\citenamefont {Kaplan}\ \emph {et~al.}(2009)\citenamefont {Kaplan},
  \citenamefont {Luty},\ and\ \citenamefont {Zurek}}]{Kaplan:2009ag}%
  \BibitemOpen
  \bibfield  {author} {\bibinfo {author} {\bibfnamefont {D.~E.}\ \bibnamefont
  {Kaplan}}, \bibinfo {author} {\bibfnamefont {M.~A.}\ \bibnamefont {Luty}}, \
  and\ \bibinfo {author} {\bibfnamefont {K.~M.}\ \bibnamefont {Zurek}},\ }\href
  {\doibase 10.1103/PhysRevD.79.115016} {\bibfield  {journal} {\bibinfo
  {journal} {Phys. Rev.}\ }\textbf {\bibinfo {volume} {D79}},\ \bibinfo {pages}
  {115016} (\bibinfo {year} {2009})},\ \Eprint {http://arxiv.org/abs/0901.4117}
  {arXiv:0901.4117 [hep-ph]} \BibitemShut {NoStop}%
\bibitem [{\citenamefont {Petraki}\ and\ \citenamefont
  {Volkas}(2013)}]{Petraki:2013wwa}%
  \BibitemOpen
  \bibfield  {author} {\bibinfo {author} {\bibfnamefont {K.}~\bibnamefont
  {Petraki}}\ and\ \bibinfo {author} {\bibfnamefont {R.~R.}\ \bibnamefont
  {Volkas}},\ }\href {\doibase 10.1142/S0217751X13300287} {\bibfield  {journal}
  {\bibinfo  {journal} {Int. J. Mod. Phys.}\ }\textbf {\bibinfo {volume}
  {A28}},\ \bibinfo {pages} {1330028} (\bibinfo {year} {2013})},\ \Eprint
  {http://arxiv.org/abs/1305.4939} {arXiv:1305.4939 [hep-ph]} \BibitemShut
  {NoStop}%
\bibitem [{\citenamefont {Zurek}(2014)}]{Zurek:2013wia}%
  \BibitemOpen
  \bibfield  {author} {\bibinfo {author} {\bibfnamefont {K.~M.}\ \bibnamefont
  {Zurek}},\ }\href {\doibase 10.1016/j.physrep.2013.12.001} {\bibfield
  {journal} {\bibinfo  {journal} {Phys. Rept.}\ }\textbf {\bibinfo {volume}
  {537}},\ \bibinfo {pages} {91} (\bibinfo {year} {2014})},\ \Eprint
  {http://arxiv.org/abs/1308.0338} {arXiv:1308.0338 [hep-ph]} \BibitemShut
  {NoStop}%
\bibitem [{\citenamefont {Knapen}\ \emph
  {et~al.}(2017{\natexlab{a}})\citenamefont {Knapen}, \citenamefont {Lin},\
  and\ \citenamefont {Zurek}}]{Knapen:2017xzo}%
  \BibitemOpen
  \bibfield  {author} {\bibinfo {author} {\bibfnamefont {S.}~\bibnamefont
  {Knapen}}, \bibinfo {author} {\bibfnamefont {T.}~\bibnamefont {Lin}}, \ and\
  \bibinfo {author} {\bibfnamefont {K.~M.}\ \bibnamefont {Zurek}},\ }\href@noop
  {} {\  (\bibinfo {year} {2017}{\natexlab{a}})},\ \Eprint
  {http://arxiv.org/abs/1709.07882} {arXiv:1709.07882 [hep-ph]} \BibitemShut
  {NoStop}%
\bibitem [{\citenamefont {Green}\ and\ \citenamefont
  {Rajendran}(2017)}]{Green:2017ybv}%
  \BibitemOpen
  \bibfield  {author} {\bibinfo {author} {\bibfnamefont {D.}~\bibnamefont
  {Green}}\ and\ \bibinfo {author} {\bibfnamefont {S.}~\bibnamefont
  {Rajendran}},\ }\href {\doibase 10.1007/JHEP10(2017)013} {\bibfield
  {journal} {\bibinfo  {journal} {JHEP}\ }\textbf {\bibinfo {volume} {10}},\
  \bibinfo {pages} {013} (\bibinfo {year} {2017})},\ \Eprint
  {http://arxiv.org/abs/1701.08750} {arXiv:1701.08750 [hep-ph]} \BibitemShut
  {NoStop}%
\bibitem [{\citenamefont {Hochberg}\ \emph
  {et~al.}(2015{\natexlab{a}})\citenamefont {Hochberg}, \citenamefont {Zhao},\
  and\ \citenamefont {Zurek}}]{Hochberg:2015pha}%
  \BibitemOpen
  \bibfield  {author} {\bibinfo {author} {\bibfnamefont {Y.}~\bibnamefont
  {Hochberg}}, \bibinfo {author} {\bibfnamefont {Y.}~\bibnamefont {Zhao}}, \
  and\ \bibinfo {author} {\bibfnamefont {K.~M.}\ \bibnamefont {Zurek}},\
  }\href@noop {} {\  (\bibinfo {year} {2015}{\natexlab{a}})},\ \Eprint
  {http://arxiv.org/abs/1504.07237} {arXiv:1504.07237 [hep-ph]} \BibitemShut
  {NoStop}%
\bibitem [{\citenamefont {Hochberg}\ \emph
  {et~al.}(2015{\natexlab{b}})\citenamefont {Hochberg}, \citenamefont {Pyle},
  \citenamefont {Zhao},\ and\ \citenamefont {Zurek}}]{Hochberg:2015fth}%
  \BibitemOpen
  \bibfield  {author} {\bibinfo {author} {\bibfnamefont {Y.}~\bibnamefont
  {Hochberg}}, \bibinfo {author} {\bibfnamefont {M.}~\bibnamefont {Pyle}},
  \bibinfo {author} {\bibfnamefont {Y.}~\bibnamefont {Zhao}}, \ and\ \bibinfo
  {author} {\bibfnamefont {K.~M.}\ \bibnamefont {Zurek}},\ }\href@noop {} {\
  (\bibinfo {year} {2015}{\natexlab{b}})},\ \Eprint
  {http://arxiv.org/abs/1512.04533} {arXiv:1512.04533 [hep-ph]} \BibitemShut
  {NoStop}%
\bibitem [{\citenamefont {Knapen}\ \emph
  {et~al.}(2017{\natexlab{b}})\citenamefont {Knapen}, \citenamefont {Lin},\
  and\ \citenamefont {Zurek}}]{Knapen:2016cue}%
  \BibitemOpen
  \bibfield  {author} {\bibinfo {author} {\bibfnamefont {S.}~\bibnamefont
  {Knapen}}, \bibinfo {author} {\bibfnamefont {T.}~\bibnamefont {Lin}}, \ and\
  \bibinfo {author} {\bibfnamefont {K.~M.}\ \bibnamefont {Zurek}},\ }\href
  {\doibase 10.1103/PhysRevD.95.056019} {\bibfield  {journal} {\bibinfo
  {journal} {Phys. Rev.}\ }\textbf {\bibinfo {volume} {D95}},\ \bibinfo {pages}
  {056019} (\bibinfo {year} {2017}{\natexlab{b}})},\ \Eprint
  {http://arxiv.org/abs/1611.06228} {arXiv:1611.06228 [hep-ph]} \BibitemShut
  {NoStop}%
\bibitem [{\citenamefont {Schutz}\ and\ \citenamefont
  {Zurek}(2016)}]{Schutz:2016tid}%
  \BibitemOpen
  \bibfield  {author} {\bibinfo {author} {\bibfnamefont {K.}~\bibnamefont
  {Schutz}}\ and\ \bibinfo {author} {\bibfnamefont {K.~M.}\ \bibnamefont
  {Zurek}},\ }\href {\doibase 10.1103/PhysRevLett.117.121302} {\bibfield
  {journal} {\bibinfo  {journal} {Phys. Rev. Lett.}\ }\textbf {\bibinfo
  {volume} {117}},\ \bibinfo {pages} {121302} (\bibinfo {year} {2016})},\
  \Eprint {http://arxiv.org/abs/1604.08206} {arXiv:1604.08206 [hep-ph]}
  \BibitemShut {NoStop}%
\bibitem [{\citenamefont {Hochberg}\ \emph
  {et~al.}(2017{\natexlab{a}})\citenamefont {Hochberg}, \citenamefont {Kahn},
  \citenamefont {Lisanti}, \citenamefont {Zurek}, \citenamefont {Grushin},
  \citenamefont {Ilan}, \citenamefont {Griffin}, \citenamefont {Liu},\ and\
  \citenamefont {Weber}}]{Hochberg:2017wce}%
  \BibitemOpen
  \bibfield  {author} {\bibinfo {author} {\bibfnamefont {Y.}~\bibnamefont
  {Hochberg}}, \bibinfo {author} {\bibfnamefont {Y.}~\bibnamefont {Kahn}},
  \bibinfo {author} {\bibfnamefont {M.}~\bibnamefont {Lisanti}}, \bibinfo
  {author} {\bibfnamefont {K.~M.}\ \bibnamefont {Zurek}}, \bibinfo {author}
  {\bibfnamefont {A.}~\bibnamefont {Grushin}}, \bibinfo {author} {\bibfnamefont
  {R.}~\bibnamefont {Ilan}}, \bibinfo {author} {\bibfnamefont {S.~M.}\
  \bibnamefont {Griffin}}, \bibinfo {author} {\bibfnamefont {Z.-F.}\
  \bibnamefont {Liu}}, \ and\ \bibinfo {author} {\bibfnamefont {S.~F.}\
  \bibnamefont {Weber}},\ }\href@noop {} {\  (\bibinfo {year}
  {2017}{\natexlab{a}})},\ \Eprint {http://arxiv.org/abs/1708.08929}
  {arXiv:1708.08929 [hep-ph]} \BibitemShut {NoStop}%
\bibitem [{\citenamefont {Griffin}\ \emph {et~al.}(tion)\citenamefont
  {Griffin}, \citenamefont {Knapen}, \citenamefont {Lin}, \citenamefont
  {Pyle},\ and\ \citenamefont {Zurek}}]{longpaper}%
  \BibitemOpen
  \bibfield  {author} {\bibinfo {author} {\bibfnamefont {S.}~\bibnamefont
  {Griffin}}, \bibinfo {author} {\bibfnamefont {S.}~\bibnamefont {Knapen}},
  \bibinfo {author} {\bibfnamefont {T.}~\bibnamefont {Lin}}, \bibinfo {author}
  {\bibfnamefont {M.}~\bibnamefont {Pyle}}, \ and\ \bibinfo {author}
  {\bibfnamefont {K.}~\bibnamefont {Zurek}},\ }\href@noop {} {} (\bibinfo
  {year} {{In preparation}})\BibitemShut {NoStop}%
\bibitem [{\citenamefont {Kittel}(2004)}]{Kittel}%
  \BibitemOpen
  \bibfield  {author} {\bibinfo {author} {\bibfnamefont {C.}~\bibnamefont
  {Kittel}},\ }\href@noop {} {\emph {\bibinfo {title} {Introduction to Solid
  State Physics}}},\ \bibinfo {edition} {8th}\ ed.\ (\bibinfo  {publisher}
  {Wiley},\ \bibinfo {year} {2004})\BibitemShut {NoStop}%
\bibitem [{\citenamefont {Giannozzi}\ and\ \citenamefont
  {et.al.}(2009)}]{0953-8984-21-39-395502}%
  \BibitemOpen
  \bibfield  {author} {\bibinfo {author} {\bibfnamefont {P.}~\bibnamefont
  {Giannozzi}}\ and\ \bibinfo {author} {\bibnamefont {et.al.}},\ }\href
  {http://stacks.iop.org/0953-8984/21/i=39/a=395502} {\bibfield  {journal}
  {\bibinfo  {journal} {Journal of Physics: Condensed Matter}\ }\textbf
  {\bibinfo {volume} {21}},\ \bibinfo {pages} {395502} (\bibinfo {year}
  {2009})}\BibitemShut {NoStop}%
\bibitem [{\citenamefont {Yu}\ and\ \citenamefont {Cardona}(2010)}]{YuCardona}%
  \BibitemOpen
  \bibfield  {author} {\bibinfo {author} {\bibfnamefont {P.~Y.}\ \bibnamefont
  {Yu}}\ and\ \bibinfo {author} {\bibfnamefont {P.}~\bibnamefont {Cardona}},\
  }\href@noop {} {\emph {\bibinfo {title} {Fundamentals of Semiconductors:
  Physics and Materials Properties}}},\ \bibinfo {edition} {4th}\ ed.\
  (\bibinfo  {publisher} {Springer},\ \bibinfo {year} {2010})\BibitemShut
  {NoStop}%
\bibitem [{\citenamefont {Lockwood}\ \emph {et~al.}(2005)\citenamefont
  {Lockwood}, \citenamefont {Yu},\ and\ \citenamefont
  {Rowell}}]{LOCKWOOD2005404}%
  \BibitemOpen
  \bibfield  {author} {\bibinfo {author} {\bibfnamefont {D.}~\bibnamefont
  {Lockwood}}, \bibinfo {author} {\bibfnamefont {G.}~\bibnamefont {Yu}}, \ and\
  \bibinfo {author} {\bibfnamefont {N.}~\bibnamefont {Rowell}},\ }\href
  {\doibase https://doi.org/10.1016/j.ssc.2005.08.030} {\bibfield  {journal}
  {\bibinfo  {journal} {Solid State Communications}\ }\textbf {\bibinfo
  {volume} {136}},\ \bibinfo {pages} {404 } (\bibinfo {year}
  {2005})}\BibitemShut {NoStop}%
\bibitem [{\citenamefont {Irmer}\ \emph {et~al.}(1996)\citenamefont {Irmer},
  \citenamefont {Wenzel},\ and\ \citenamefont {Monecke}}]{PSSB:PSSB2221950110}%
  \BibitemOpen
  \bibfield  {author} {\bibinfo {author} {\bibfnamefont {G.}~\bibnamefont
  {Irmer}}, \bibinfo {author} {\bibfnamefont {M.}~\bibnamefont {Wenzel}}, \
  and\ \bibinfo {author} {\bibfnamefont {J.}~\bibnamefont {Monecke}},\ }\href
  {\doibase 10.1002/pssb.2221950110} {\bibfield  {journal} {\bibinfo  {journal}
  {physica status solidi (b)}\ }\textbf {\bibinfo {volume} {195}},\ \bibinfo
  {pages} {85} (\bibinfo {year} {1996})}\BibitemShut {NoStop}%
\bibitem [{\citenamefont {{Pyle}}\ \emph {et~al.}(2015)\citenamefont {{Pyle}},
  \citenamefont {{Figueroa-Feliciano}},\ and\ \citenamefont
  {{Sadoulet}}}]{Pyle_15_OptimizedCalorimeters}%
  \BibitemOpen
  \bibfield  {author} {\bibinfo {author} {\bibfnamefont {M.}~\bibnamefont
  {{Pyle}}}, \bibinfo {author} {\bibfnamefont {E.}~\bibnamefont
  {{Figueroa-Feliciano}}}, \ and\ \bibinfo {author} {\bibfnamefont
  {B.}~\bibnamefont {{Sadoulet}}},\ }\href@noop {} {\bibfield  {journal}
  {\bibinfo  {journal} {ArXiv e-prints}\ } (\bibinfo {year} {2015})},\ \Eprint
  {http://arxiv.org/abs/1503.01200} {arXiv:1503.01200 [astro-ph.IM]}
  \BibitemShut {NoStop}%
\bibitem [{\citenamefont {Sinvani}\ \emph {et~al.}(1983)\citenamefont
  {Sinvani}, \citenamefont {Taborek},\ and\ \citenamefont
  {Goodstein}}]{Sivani_95PLA_HeAdsorptionCrystal}%
  \BibitemOpen
  \bibfield  {author} {\bibinfo {author} {\bibfnamefont {M.}~\bibnamefont
  {Sinvani}}, \bibinfo {author} {\bibfnamefont {P.}~\bibnamefont {Taborek}}, \
  and\ \bibinfo {author} {\bibfnamefont {D.}~\bibnamefont {Goodstein}},\ }\href
  {\doibase https://doi.org/10.1016/0375-9601(83)90782-X} {\bibfield  {journal}
  {\bibinfo  {journal} {Physics Letters A}\ }\textbf {\bibinfo {volume} {95}},\
  \bibinfo {pages} {59 } (\bibinfo {year} {1983})}\BibitemShut {NoStop}%
\bibitem [{\citenamefont {More}\ \emph {et~al.}(1996)\citenamefont {More},
  \citenamefont {Adams}, \citenamefont {Bandler}, \citenamefont {Brou\"er},
  \citenamefont {Lanou}, \citenamefont {Maris},\ and\ \citenamefont
  {Seidel}}]{More_96PRB_PhononAmplifier}%
  \BibitemOpen
  \bibfield  {author} {\bibinfo {author} {\bibfnamefont {T.}~\bibnamefont
  {More}}, \bibinfo {author} {\bibfnamefont {J.~S.}\ \bibnamefont {Adams}},
  \bibinfo {author} {\bibfnamefont {S.~R.}\ \bibnamefont {Bandler}}, \bibinfo
  {author} {\bibfnamefont {S.~M.}\ \bibnamefont {Brou\"er}}, \bibinfo {author}
  {\bibfnamefont {R.~E.}\ \bibnamefont {Lanou}}, \bibinfo {author}
  {\bibfnamefont {H.~J.}\ \bibnamefont {Maris}}, \ and\ \bibinfo {author}
  {\bibfnamefont {G.~M.}\ \bibnamefont {Seidel}},\ }\href {\doibase
  10.1103/PhysRevB.54.534} {\bibfield  {journal} {\bibinfo  {journal} {Phys.
  Rev. B}\ }\textbf {\bibinfo {volume} {54}},\ \bibinfo {pages} {534} (\bibinfo
  {year} {1996})}\BibitemShut {NoStop}%
\bibitem [{\citenamefont {Maris}\ \emph {et~al.}(2017)\citenamefont {Maris},
  \citenamefont {Seidel},\ and\ \citenamefont
  {Stein}}]{Maris_17PRL_HeIonization}%
  \BibitemOpen
  \bibfield  {author} {\bibinfo {author} {\bibfnamefont {H.~J.}\ \bibnamefont
  {Maris}}, \bibinfo {author} {\bibfnamefont {G.~M.}\ \bibnamefont {Seidel}}, \
  and\ \bibinfo {author} {\bibfnamefont {D.}~\bibnamefont {Stein}},\ }\href
  {\doibase 10.1103/PhysRevLett.119.181303} {\bibfield  {journal} {\bibinfo
  {journal} {Phys. Rev. Lett.}\ }\textbf {\bibinfo {volume} {119}},\ \bibinfo
  {pages} {181303} (\bibinfo {year} {2017})}\BibitemShut {NoStop}%
\bibitem [{\citenamefont {Frank}\ \emph {et~al.}(1994)\citenamefont {Frank},
  \citenamefont {Dummer}, \citenamefont {Cooper}, \citenamefont {Igalson},
  \citenamefont {Probst},\ and\ \citenamefont {Seidel}}]{FRANK1994367}%
  \BibitemOpen
  \bibfield  {author} {\bibinfo {author} {\bibfnamefont {M.}~\bibnamefont
  {Frank}}, \bibinfo {author} {\bibfnamefont {D.}~\bibnamefont {Dummer}},
  \bibinfo {author} {\bibfnamefont {S.}~\bibnamefont {Cooper}}, \bibinfo
  {author} {\bibfnamefont {J.}~\bibnamefont {Igalson}}, \bibinfo {author}
  {\bibfnamefont {F.}~\bibnamefont {Probst}}, \ and\ \bibinfo {author}
  {\bibfnamefont {W.}~\bibnamefont {Seidel}},\ }\href {\doibase
  https://doi.org/10.1016/0168-9002(94)91015-4} {\bibfield  {journal} {\bibinfo
   {journal} {Nuclear Instruments and Methods in Physics Research Section A:
  Accelerators, Spectrometers, Detectors and Associated Equipment}\ }\textbf
  {\bibinfo {volume} {345}},\ \bibinfo {pages} {367 } (\bibinfo {year}
  {1994})}\BibitemShut {NoStop}%
\bibitem [{\citenamefont {Irwin}\ \emph {et~al.}(1995)\citenamefont {Irwin},
  \citenamefont {Nam}, \citenamefont {Cabrera}, \citenamefont {Chugg},\ and\
  \citenamefont {Young}}]{Irwin_95RSI_QET}%
  \BibitemOpen
  \bibfield  {author} {\bibinfo {author} {\bibfnamefont {K.~D.}\ \bibnamefont
  {Irwin}}, \bibinfo {author} {\bibfnamefont {S.~W.}\ \bibnamefont {Nam}},
  \bibinfo {author} {\bibfnamefont {B.}~\bibnamefont {Cabrera}}, \bibinfo
  {author} {\bibfnamefont {B.}~\bibnamefont {Chugg}}, \ and\ \bibinfo {author}
  {\bibfnamefont {B.~A.}\ \bibnamefont {Young}},\ }\href {\doibase
  10.1063/1.1146105} {\bibfield  {journal} {\bibinfo  {journal} {Review of
  Scientific Instruments}\ }\textbf {\bibinfo {volume} {66}},\ \bibinfo {pages}
  {5322} (\bibinfo {year} {1995})},\ \Eprint
  {http://arxiv.org/abs/https://doi.org/10.1063/1.1146105}
  {https://doi.org/10.1063/1.1146105} \BibitemShut {NoStop}%
\bibitem [{\citenamefont {Tamura}(1985)}]{Tamura_85PRB_AHDC}%
  \BibitemOpen
  \bibfield  {author} {\bibinfo {author} {\bibfnamefont {S.-i.}\ \bibnamefont
  {Tamura}},\ }\href {\doibase 10.1103/PhysRevB.31.2574} {\bibfield  {journal}
  {\bibinfo  {journal} {Phys. Rev. B}\ }\textbf {\bibinfo {volume} {31}},\
  \bibinfo {pages} {2574} (\bibinfo {year} {1985})}\BibitemShut {NoStop}%
\bibitem [{\citenamefont {Knaak}\ \emph {et~al.}(1986)\citenamefont {Knaak},
  \citenamefont {Hau{\ss}}, \citenamefont {Kummrow},\ and\ \citenamefont
  {Mei{\ss}ner}}]{Knaak_86_SurfaceThermalization}%
  \BibitemOpen
  \bibfield  {author} {\bibinfo {author} {\bibfnamefont {W.}~\bibnamefont
  {Knaak}}, \bibinfo {author} {\bibfnamefont {T.}~\bibnamefont {Hau{\ss}}},
  \bibinfo {author} {\bibfnamefont {M.}~\bibnamefont {Kummrow}}, \ and\
  \bibinfo {author} {\bibfnamefont {M.}~\bibnamefont {Mei{\ss}ner}},\ }\enquote
  {\bibinfo {title} {Thermalization of ballistic phonon pulses in dielectric
  crystals below 1k using time-resolved thermometry},}\ in\ \href {\doibase
  10.1007/978-3-642-82912-3_52} {\emph {\bibinfo {booktitle} {Phonon Scattering
  in Condensed Matter V: Proceedings of the Fifth International Conference
  Urbana, Illinois, June 2--6, 1986}}},\ \bibinfo {editor} {edited by\ \bibinfo
  {editor} {\bibfnamefont {A.~C.}\ \bibnamefont {Anderson}}\ and\ \bibinfo
  {editor} {\bibfnamefont {J.~P.}\ \bibnamefont {Wolfe}}}\ (\bibinfo
  {publisher} {Springer Berlin Heidelberg},\ \bibinfo {address} {Berlin,
  Heidelberg},\ \bibinfo {year} {1986})\ pp.\ \bibinfo {pages}
  {174--176}\BibitemShut {NoStop}%
\bibitem [{\citenamefont {Agnese}\ \emph
  {et~al.}(2017{\natexlab{b}})\citenamefont {Agnese} \emph
  {et~al.}}]{SCDMS_17PRD_SensitivityProjections}%
  \BibitemOpen
  \bibfield  {author} {\bibinfo {author} {\bibfnamefont {R.}~\bibnamefont
  {Agnese}} \emph {et~al.} (\bibinfo {collaboration} {SuperCDMS
  Collaboration}),\ }\href {\doibase 10.1103/PhysRevD.95.082002} {\bibfield
  {journal} {\bibinfo  {journal} {Phys. Rev. D}\ }\textbf {\bibinfo {volume}
  {95}},\ \bibinfo {pages} {082002} (\bibinfo {year}
  {2017}{\natexlab{b}})}\BibitemShut {NoStop}%
\bibitem [{\citenamefont {{Robinson}}(2017)}]{2016arXiv161007656R}%
  \BibitemOpen
  \bibfield  {author} {\bibinfo {author} {\bibfnamefont {A.~E.}\ \bibnamefont
  {{Robinson}}},\ }\href {\doibase 10.1103/PhysRevD.95.021301} {\bibfield
  {journal} {\bibinfo  {journal} {\prd}\ }\textbf {\bibinfo {volume} {95}},\
  \bibinfo {eid} {021301} (\bibinfo {year} {2017})},\ \Eprint
  {http://arxiv.org/abs/1610.07656} {arXiv:1610.07656 [astro-ph.IM]}
  \BibitemShut {NoStop}%
\bibitem [{\citenamefont {An}\ \emph {et~al.}(2015)\citenamefont {An},
  \citenamefont {Pospelov}, \citenamefont {Pradler},\ and\ \citenamefont
  {Ritz}}]{An:2014twa}%
  \BibitemOpen
  \bibfield  {author} {\bibinfo {author} {\bibfnamefont {H.}~\bibnamefont
  {An}}, \bibinfo {author} {\bibfnamefont {M.}~\bibnamefont {Pospelov}},
  \bibinfo {author} {\bibfnamefont {J.}~\bibnamefont {Pradler}}, \ and\
  \bibinfo {author} {\bibfnamefont {A.}~\bibnamefont {Ritz}},\ }\href {\doibase
  10.1016/j.physletb.2015.06.018} {\bibfield  {journal} {\bibinfo  {journal}
  {Phys. Lett.}\ }\textbf {\bibinfo {volume} {B747}},\ \bibinfo {pages} {331}
  (\bibinfo {year} {2015})},\ \Eprint {http://arxiv.org/abs/1412.8378}
  {arXiv:1412.8378 [hep-ph]} \BibitemShut {NoStop}%
\bibitem [{\citenamefont {Hochberg}\ \emph
  {et~al.}(2016{\natexlab{b}})\citenamefont {Hochberg}, \citenamefont {Lin},\
  and\ \citenamefont {Zurek}}]{Hochberg:2016ajh}%
  \BibitemOpen
  \bibfield  {author} {\bibinfo {author} {\bibfnamefont {Y.}~\bibnamefont
  {Hochberg}}, \bibinfo {author} {\bibfnamefont {T.}~\bibnamefont {Lin}}, \
  and\ \bibinfo {author} {\bibfnamefont {K.~M.}\ \bibnamefont {Zurek}},\ }\href
  {\doibase 10.1103/PhysRevD.94.015019} {\bibfield  {journal} {\bibinfo
  {journal} {Phys. Rev.}\ }\textbf {\bibinfo {volume} {D94}},\ \bibinfo {pages}
  {015019} (\bibinfo {year} {2016}{\natexlab{b}})},\ \Eprint
  {http://arxiv.org/abs/1604.06800} {arXiv:1604.06800 [hep-ph]} \BibitemShut
  {NoStop}%
\bibitem [{\citenamefont {An}\ \emph {et~al.}(2013{\natexlab{a}})\citenamefont
  {An}, \citenamefont {Pospelov},\ and\ \citenamefont {Pradler}}]{An:2013yfc}%
  \BibitemOpen
  \bibfield  {author} {\bibinfo {author} {\bibfnamefont {H.}~\bibnamefont
  {An}}, \bibinfo {author} {\bibfnamefont {M.}~\bibnamefont {Pospelov}}, \ and\
  \bibinfo {author} {\bibfnamefont {J.}~\bibnamefont {Pradler}},\ }\href
  {\doibase 10.1016/j.physletb.2013.07.008} {\bibfield  {journal} {\bibinfo
  {journal} {Phys. Lett.}\ }\textbf {\bibinfo {volume} {B725}},\ \bibinfo
  {pages} {190} (\bibinfo {year} {2013}{\natexlab{a}})},\ \Eprint
  {http://arxiv.org/abs/1302.3884} {arXiv:1302.3884 [hep-ph]} \BibitemShut
  {NoStop}%
\bibitem [{\citenamefont {An}\ \emph {et~al.}(2013{\natexlab{b}})\citenamefont
  {An}, \citenamefont {Pospelov},\ and\ \citenamefont {Pradler}}]{An:2013yua}%
  \BibitemOpen
  \bibfield  {author} {\bibinfo {author} {\bibfnamefont {H.}~\bibnamefont
  {An}}, \bibinfo {author} {\bibfnamefont {M.}~\bibnamefont {Pospelov}}, \ and\
  \bibinfo {author} {\bibfnamefont {J.}~\bibnamefont {Pradler}},\ }\href
  {\doibase 10.1103/PhysRevLett.111.041302} {\bibfield  {journal} {\bibinfo
  {journal} {Phys. Rev. Lett.}\ }\textbf {\bibinfo {volume} {111}},\ \bibinfo
  {pages} {041302} (\bibinfo {year} {2013}{\natexlab{b}})},\ \Eprint
  {http://arxiv.org/abs/1304.3461} {arXiv:1304.3461 [hep-ph]} \BibitemShut
  {NoStop}%
\bibitem [{\citenamefont {Aguilar-Arevalo}\ \emph {et~al.}(2017)\citenamefont
  {Aguilar-Arevalo} \emph {et~al.}}]{Aguilar-Arevalo:2016zop}%
  \BibitemOpen
  \bibfield  {author} {\bibinfo {author} {\bibfnamefont {A.}~\bibnamefont
  {Aguilar-Arevalo}} \emph {et~al.} (\bibinfo {collaboration} {DAMIC}),\ }\href
  {\doibase 10.1103/PhysRevLett.118.141803} {\bibfield  {journal} {\bibinfo
  {journal} {Phys. Rev. Lett.}\ }\textbf {\bibinfo {volume} {118}},\ \bibinfo
  {pages} {141803} (\bibinfo {year} {2017})},\ \Eprint
  {http://arxiv.org/abs/1611.03066} {arXiv:1611.03066 [astro-ph.CO]}
  \BibitemShut {NoStop}%
\bibitem [{\citenamefont {Bloch}\ \emph {et~al.}(2017)\citenamefont {Bloch},
  \citenamefont {Essig}, \citenamefont {Tobioka}, \citenamefont {Volansky},\
  and\ \citenamefont {Yu}}]{Bloch:2016sjj}%
  \BibitemOpen
  \bibfield  {author} {\bibinfo {author} {\bibfnamefont {I.~M.}\ \bibnamefont
  {Bloch}}, \bibinfo {author} {\bibfnamefont {R.}~\bibnamefont {Essig}},
  \bibinfo {author} {\bibfnamefont {K.}~\bibnamefont {Tobioka}}, \bibinfo
  {author} {\bibfnamefont {T.}~\bibnamefont {Volansky}}, \ and\ \bibinfo
  {author} {\bibfnamefont {T.-T.}\ \bibnamefont {Yu}},\ }\href {\doibase
  10.1007/JHEP06(2017)087} {\bibfield  {journal} {\bibinfo  {journal} {JHEP}\
  }\textbf {\bibinfo {volume} {06}},\ \bibinfo {pages} {087} (\bibinfo {year}
  {2017})},\ \Eprint {http://arxiv.org/abs/1608.02123} {arXiv:1608.02123
  [hep-ph]} \BibitemShut {NoStop}%
\bibitem [{\citenamefont {Hochberg}\ \emph
  {et~al.}(2017{\natexlab{b}})\citenamefont {Hochberg}, \citenamefont {Lin},\
  and\ \citenamefont {Zurek}}]{Hochberg:2016sqx}%
  \BibitemOpen
  \bibfield  {author} {\bibinfo {author} {\bibfnamefont {Y.}~\bibnamefont
  {Hochberg}}, \bibinfo {author} {\bibfnamefont {T.}~\bibnamefont {Lin}}, \
  and\ \bibinfo {author} {\bibfnamefont {K.~M.}\ \bibnamefont {Zurek}},\ }\href
  {\doibase 10.1103/PhysRevD.95.023013} {\bibfield  {journal} {\bibinfo
  {journal} {Phys. Rev.}\ }\textbf {\bibinfo {volume} {D95}},\ \bibinfo {pages}
  {023013} (\bibinfo {year} {2017}{\natexlab{b}})},\ \Eprint
  {http://arxiv.org/abs/1608.01994} {arXiv:1608.01994 [hep-ph]} \BibitemShut
  {NoStop}%
\bibitem [{\citenamefont {Arvanitaki}\ \emph {et~al.}(2017)\citenamefont
  {Arvanitaki}, \citenamefont {Dimopoulos},\ and\ \citenamefont
  {Van~Tilburg}}]{Arvanitaki:2017nhi}%
  \BibitemOpen
  \bibfield  {author} {\bibinfo {author} {\bibfnamefont {A.}~\bibnamefont
  {Arvanitaki}}, \bibinfo {author} {\bibfnamefont {S.}~\bibnamefont
  {Dimopoulos}}, \ and\ \bibinfo {author} {\bibfnamefont {K.}~\bibnamefont
  {Van~Tilburg}},\ }\href@noop {} {\  (\bibinfo {year} {2017})},\ \Eprint
  {http://arxiv.org/abs/1709.05354} {arXiv:1709.05354 [hep-ph]} \BibitemShut
  {NoStop}%
\bibitem [{\citenamefont {Bunting}\ \emph {et~al.}(2017)\citenamefont
  {Bunting}, \citenamefont {Gratta}, \citenamefont {Melia},\ and\ \citenamefont
  {Rajendran}}]{Bunting:2017net}%
  \BibitemOpen
  \bibfield  {author} {\bibinfo {author} {\bibfnamefont {P.~C.}\ \bibnamefont
  {Bunting}}, \bibinfo {author} {\bibfnamefont {G.}~\bibnamefont {Gratta}},
  \bibinfo {author} {\bibfnamefont {T.}~\bibnamefont {Melia}}, \ and\ \bibinfo
  {author} {\bibfnamefont {S.}~\bibnamefont {Rajendran}},\ }\href {\doibase
  10.1103/PhysRevD.95.095001} {\bibfield  {journal} {\bibinfo  {journal} {Phys.
  Rev.}\ }\textbf {\bibinfo {volume} {D95}},\ \bibinfo {pages} {095001}
  (\bibinfo {year} {2017})},\ \Eprint {http://arxiv.org/abs/1701.06566}
  {arXiv:1701.06566 [hep-ph]} \BibitemShut {NoStop}%
\bibitem [{\citenamefont {Lawler}\ and\ \citenamefont
  {Shirley}(2004)}]{PhysRevB.70.245209}%
  \BibitemOpen
  \bibfield  {author} {\bibinfo {author} {\bibfnamefont {H.~M.}\ \bibnamefont
  {Lawler}}\ and\ \bibinfo {author} {\bibfnamefont {E.~L.}\ \bibnamefont
  {Shirley}},\ }\href {\doibase 10.1103/PhysRevB.70.245209} {\bibfield
  {journal} {\bibinfo  {journal} {Phys. Rev. B}\ }\textbf {\bibinfo {volume}
  {70}},\ \bibinfo {pages} {245209} (\bibinfo {year} {2004})}\BibitemShut
  {NoStop}%
\bibitem [{\citenamefont {Palik}(1985)}]{opticalconstantsGaAs}%
  \BibitemOpen
  \bibfield  {author} {\bibinfo {author} {\bibfnamefont {E.~D.}\ \bibnamefont
  {Palik}},\ }in\ \href@noop {} {\emph {\bibinfo {booktitle} {Handbook of
  Optical Constants of Solids}}},\ \bibinfo {editor} {edited by\ \bibinfo
  {editor} {\bibfnamefont {E.~D.}\ \bibnamefont {Palik}}}\ (\bibinfo
  {publisher} {Elsevier},\ \bibinfo {year} {1985})\ pp.\ \bibinfo {pages}
  {429--443}\BibitemShut {NoStop}%
\bibitem [{\citenamefont {Fr{\"o}hlich}(1954)}]{frolich1954}%
  \BibitemOpen
  \bibfield  {author} {\bibinfo {author} {\bibfnamefont {H.}~\bibnamefont
  {Fr{\"o}hlich}},\ }\href {\doibase 10.1080/00018735400101213} {\bibfield
  {journal} {\bibinfo  {journal} {Advances in Physics}\ }\textbf {\bibinfo
  {volume} {3}},\ \bibinfo {pages} {325} (\bibinfo {year} {1954})},\ \Eprint
  {http://arxiv.org/abs/https://doi.org/10.1080/00018735400101213}
  {https://doi.org/10.1080/00018735400101213} \BibitemShut {NoStop}%
\bibitem [{\citenamefont {Born}\ and\ \citenamefont {Huang}(1954)}]{BornHuang}%
  \BibitemOpen
  \bibfield  {author} {\bibinfo {author} {\bibfnamefont {M.}~\bibnamefont
  {Born}}\ and\ \bibinfo {author} {\bibfnamefont {K.}~\bibnamefont {Huang}},\
  }\href@noop {} {\emph {\bibinfo {title} {Dynamical Theory of Crystal
  Lattices}}}\ (\bibinfo  {publisher} {Clarendon Press},\ \bibinfo {year}
  {1954})\BibitemShut {NoStop}%
\bibitem [{\citenamefont {Verdi}\ and\ \citenamefont
  {Giustino}(2015)}]{PhysRevLett.115.176401}%
  \BibitemOpen
  \bibfield  {author} {\bibinfo {author} {\bibfnamefont {C.}~\bibnamefont
  {Verdi}}\ and\ \bibinfo {author} {\bibfnamefont {F.}~\bibnamefont
  {Giustino}},\ }\href {\doibase 10.1103/PhysRevLett.115.176401} {\bibfield
  {journal} {\bibinfo  {journal} {Phys. Rev. Lett.}\ }\textbf {\bibinfo
  {volume} {115}},\ \bibinfo {pages} {176401} (\bibinfo {year}
  {2015})}\BibitemShut {NoStop}%
\bibitem [{\citenamefont {Zhou}\ and\ \citenamefont
  {Bernardi}(2016)}]{PhysRevB.94.201201}%
  \BibitemOpen
  \bibfield  {author} {\bibinfo {author} {\bibfnamefont {J.-J.}\ \bibnamefont
  {Zhou}}\ and\ \bibinfo {author} {\bibfnamefont {M.}~\bibnamefont
  {Bernardi}},\ }\href {\doibase 10.1103/PhysRevB.94.201201} {\bibfield
  {journal} {\bibinfo  {journal} {Phys. Rev. B}\ }\textbf {\bibinfo {volume}
  {94}},\ \bibinfo {pages} {201201} (\bibinfo {year} {2016})}\BibitemShut
  {NoStop}%
\bibitem [{\citenamefont {Sjakste}\ \emph {et~al.}(2015)\citenamefont
  {Sjakste}, \citenamefont {Vast}, \citenamefont {Calandra},\ and\
  \citenamefont {Mauri}}]{PhysRevB.92.054307}%
  \BibitemOpen
  \bibfield  {author} {\bibinfo {author} {\bibfnamefont {J.}~\bibnamefont
  {Sjakste}}, \bibinfo {author} {\bibfnamefont {N.}~\bibnamefont {Vast}},
  \bibinfo {author} {\bibfnamefont {M.}~\bibnamefont {Calandra}}, \ and\
  \bibinfo {author} {\bibfnamefont {F.}~\bibnamefont {Mauri}},\ }\href
  {\doibase 10.1103/PhysRevB.92.054307} {\bibfield  {journal} {\bibinfo
  {journal} {Phys. Rev. B}\ }\textbf {\bibinfo {volume} {92}},\ \bibinfo
  {pages} {054307} (\bibinfo {year} {2015})}\BibitemShut {NoStop}%
\bibitem [{\citenamefont {Vogel}\ and\ \citenamefont
  {Redondo}(2014)}]{Vogel:2013raa}%
  \BibitemOpen
  \bibfield  {author} {\bibinfo {author} {\bibfnamefont {H.}~\bibnamefont
  {Vogel}}\ and\ \bibinfo {author} {\bibfnamefont {J.}~\bibnamefont
  {Redondo}},\ }\href {\doibase 10.1088/1475-7516/2014/02/029} {\bibfield
  {journal} {\bibinfo  {journal} {JCAP}\ }\textbf {\bibinfo {volume} {1402}},\
  \bibinfo {pages} {029} (\bibinfo {year} {2014})},\ \Eprint
  {http://arxiv.org/abs/1311.2600} {arXiv:1311.2600 [hep-ph]} \BibitemShut
  {NoStop}%
\bibitem [{\citenamefont {Davidson}\ \emph {et~al.}(2000)\citenamefont
  {Davidson}, \citenamefont {Hannestad},\ and\ \citenamefont
  {Raffelt}}]{Davidson:2000hf}%
  \BibitemOpen
  \bibfield  {author} {\bibinfo {author} {\bibfnamefont {S.}~\bibnamefont
  {Davidson}}, \bibinfo {author} {\bibfnamefont {S.}~\bibnamefont {Hannestad}},
  \ and\ \bibinfo {author} {\bibfnamefont {G.}~\bibnamefont {Raffelt}},\ }\href
  {\doibase 10.1088/1126-6708/2000/05/003} {\bibfield  {journal} {\bibinfo
  {journal} {JHEP}\ }\textbf {\bibinfo {volume} {05}},\ \bibinfo {pages} {003}
  (\bibinfo {year} {2000})},\ \Eprint {http://arxiv.org/abs/hep-ph/0001179}
  {arXiv:hep-ph/0001179 [hep-ph]} \BibitemShut {NoStop}%
\bibitem [{\citenamefont {Battaglieri}\ \emph {et~al.}(2017)\citenamefont
  {Battaglieri} \emph {et~al.}}]{Battaglieri:2017aum}%
  \BibitemOpen
  \bibfield  {author} {\bibinfo {author} {\bibfnamefont {M.}~\bibnamefont
  {Battaglieri}} \emph {et~al.},\ }\href@noop {} {\  (\bibinfo {year}
  {2017})},\ \Eprint {http://arxiv.org/abs/1707.04591} {arXiv:1707.04591
  [hep-ph]} \BibitemShut {NoStop}%
\bibitem [{\citenamefont {Lewin}\ and\ \citenamefont
  {Smith}(1996)}]{Lewin:1995rx}%
  \BibitemOpen
  \bibfield  {author} {\bibinfo {author} {\bibfnamefont {J.~D.}\ \bibnamefont
  {Lewin}}\ and\ \bibinfo {author} {\bibfnamefont {P.~F.}\ \bibnamefont
  {Smith}},\ }\href {\doibase 10.1016/S0927-6505(96)00047-3} {\bibfield
  {journal} {\bibinfo  {journal} {Astropart. Phys.}\ }\textbf {\bibinfo
  {volume} {6}},\ \bibinfo {pages} {87} (\bibinfo {year} {1996})}\BibitemShut
  {NoStop}%
\bibitem [{\citenamefont {Schober}(2014)}]{Schober2014}%
  \BibitemOpen
  \bibfield  {author} {\bibinfo {author} {\bibfnamefont {H.}~\bibnamefont
  {Schober}},\ }\href@noop {} {\bibfield  {journal} {\bibinfo  {journal}
  {Journal of Neutron Research}\ }\textbf {\bibinfo {volume} {17}},\ \bibinfo
  {pages} {109} (\bibinfo {year} {2014})}\BibitemShut {NoStop}%
\end{thebibliography}%

\end{document}